\documentclass[11pt]{article} 
\usepackage{times} 
\usepackage{graphicx} 

\begin{document}

\title{Anomalous velocity distributions in inelastic Maxwell gases}            

\author{R. Brito\\
Depto de Fisica Aplicada I and GISC, Universidad
Complutense\\ 28040 Madrid, Spain\\
Email: {\tt brito@seneca.fis.ucm.es}
\and M. H. Ernst  \\
Instituut voor Theoretische Fysica, Universiteit Utrecht\\Postbus
80.195, 3508 TD Utrecht, The Netherlands\\
Email: {\tt M.H.Ernst@phys.uu.nl}}



\date{}
\maketitle                            

\newcommand{\Label}[1]{\label{#1}}                  
\newcommand{\Bibitem}[1]{\bibitem{#1}}  

\newcommand{\largersim}{\raisebox{-.3ex}{$\enskip\stackrel{>}{\scriptstyle\sim}\enskip$}}
\newcommand{\smallersim}{\raisebox{-.3ex}{$\enskip\stackrel{<}{\scriptstyle\sim}\enskip$}}

\newcommand\be{\begin{equation}}
\newcommand\ee{\end{equation}}
\newcommand\ba{\begin{eqnarray}}
\newcommand\ea{\end{eqnarray}}
\newcommand{\nn}{\nonumber\\}
\newcommand{\av}[1]{\langle #1\rangle}  
\newcommand{\inprod}[2]{\langle #1 | #2 \rangle} 
\newcommand{\half}{\textstyle{\frac{1}{2}}}
\newcommand{\third}{\textstyle{\frac{1}{3}}}
\newcommand{\fourth}{\textstyle{\frac{1}{4}}}
\newcommand{\bhat}[1]{\hat{{\mbox{\boldmath$#1$}}}}
\newcommand{\tbkmin}{\tilde{{\bf k}}_-}
\newcommand{\tbkplus}{\tilde{{\bf k}}_+}
\newcommand{\tkmin}{\tilde{{k}}_-}
\newcommand{\gpar}{g_\parallel}
\newcommand{\bfa}{{\mbox{\boldmath$a$}}}
\newcommand{\bk}{{\mbox{\boldmath$k$}}}
\newcommand{\bq}{{\mbox{\boldmath$q$}}}
\newcommand{\bn}{{\mbox{\boldmath$n$}}}
\newcommand{\bg}{{\mbox{\boldmath$g$}}}
\newcommand{\bG}{{\mbox{\boldmath$G$}}}
\newcommand{\bv}{{\mbox{\boldmath$v$}}}
\newcommand{\bc}{{\mbox{\boldmath$c$}}}
\newcommand{\bw}{{\mbox{\boldmath$w$}}}
\newcommand{\br}{{\mbox{\boldmath$r$}}}
\newcommand{\bfu}{{\mbox{\boldmath$u$}}}
\newcommand{\bnabla} {{\mbox{\boldmath$\nabla$}}}
\newcommand{\xovery}[2]{\left( \begin{array}{c} #1 \\ #2
\end{array} \right)}
\newcommand{\eps}{{\varepsilon}}
\newcommand{\Ref}[1]{(\ref{#1})}
\newcommand{\bu}{$\bullet$}
\newcommand{\combin}[2]{\left( \begin{array}{c} #1 \\
                        #2 \end{array} \right)}
\newcommand{\tphi}{\tilde\phi}
\renewcommand{\theequation}{\arabic{section}.\arabic{equation}}

This  review is a kinetic theory study investigating the effects of
inelasticity on the structure of the non-equilibrium states, in
particular on the behavior of the velocity distribution in the high
energy tails. Starting point is the nonlinear Boltzmann equation for
spatially homogeneous systems, which supposedly describes the behavior
of the velocity distribution function in dissipative systems as long
as the system remains in the homogeneous cooling state, i.e. on
relatively short time scales before the clustering and similar
instabilities start to create spatial inhomogeneities. This is done
for the two most common models for dissipative systems, i.e. inelastic
hard spheres and inelastic Maxwell particles. There is a strong
emphasis on the latter models because that is the area where most of
the interesting new developments occurred. In systems of Maxwell
particles the collision frequency is independent of the relative
velocity of the colliding particles, and in hard sphere systems it is
linear. We then demonstrate the existence of scaling solutions for the
velocity distribution function, $F(v,t) \sim v_0(t)^{-d}
f((v/v_0(t))$, where $v_0$ is the r.m.s.~velocity. The scaling form
$f(c)$ shows overpopulation in the high energy tails. In the case of
freely cooling systems the tails are of algebraic form, $ f(c )\sim
c^{-d-a}$, where the exponent  $a$ may or may not depend on the degree
of inelasticity, and in the case of forced systems the tails are of
stretched Gaussian type
 $f(v)\sim\exp[-\beta (v/v_0)^b]$ with $b <2$.


\setcounter{section}{0}
\setcounter{subsection}{0}\setcounter{equation}{0}
\section{Introduction}

The interest in granular fluids \cite{jaeger} and gases has led to a
great revival in kinetic theory of dissipative systems
\cite{jenkins,goldhirsch,springer1,springer2}, in particular in the
non-equilibrium steady states of such systems. A granular fluid
\cite{campbell} is a collection of small or large macroscopic
particles with short range hard core repulsions, which lose energy in
inelastic collisions, and the system cools without constant energy
input.

As energy is not conserved in inelastic collisions Gibbs' equilibrium
statistical mechanics is not applicable, and non-equilibrium
statistical mechanics and kinetic theory for such systems have to be
developed to describe and  understand the wealth of interesting
phenomena discovered in such systems. The inelasticity is responsible
for a lot of new physics, such as clustering and spatial
heterogeneities \cite{G+Z}, inelastic collapse and the development of
singularities within a finite time \cite{McNamara,Kadanof},
spontaneous formations of patterns and phase transitions \cite{melo},
overpopulated non-Gaussian high energy tails in distribution functions
\cite{esipov,vNE98,experim}, break down of molecular chaos
\cite{G+Z,ignacio-II}, single peak initial distributions developing
into stable two-peak distributions as the inelasticity decreases, at
least in one-dimensional systems \cite{BBRTW,pulverenti}. These
phenomena have been studied in laboratory experiments \cite{experim},
by  Molecular Dynamics \cite{G+Z,ignacio-II,BBRTW} or Monte Carlo
simulations \cite{Brey1,MS00,BBRTW,orsay}, and by kinetic theory
methods (see recent review \cite{dufty-review} and references
therein).

The prototypical model for granular fluids or gases is a system of
mono-dis\-perse, smooth inelastic hard spheres, which lose a fraction
of their relative kinetic energy in every collision, proportional to
the degree of inelasticity $(1- \alpha^2)$, where $\alpha$ with $ 0
\leq \alpha  \leq 1$ is the coefficient of restitution. The model is a
well-defined  microscopic $N$ particle model, which can be studied by
molecular dynamics, and by kinetic theory. The single particle
distribution function can be described by the nonlinear Boltzmann
equation for inelastic hard spheres \cite{jenkins,goldhirsch}.

The present article presents a review of kinetic theory studies,
dealing with the {\it early} stages of local relaxation of the
velocity distribution $F(v,t)$, and we avoid the long time
hydrodynamic regime where gradients in density and flow fields are
important. So, we restrict our study  to {\it spatially homogeneous
states}. Without energy supply these systems are freely cooling
\cite{BDS,esipov,vNE98}.  When energy is supplied to the system a
source or forcing term is added to the Boltzmann equation
\cite{williams+mackint,ignacio-II,vNE98,MS00}, and the kinetic
equation allows steady state solutions, which depend on the mode of
energy supply \cite{williams+mackint,ignacio-II,vNE98,MS00}.

A freely evolving inelastic gas or fluid relaxes within a mean free
time to a homogeneous cooling state, where it can be described by a
scaling or similarity solution of the Boltzmann equation, $F(v,t) \sim
(1/v_0(t))^d f(v/v_0(t))$. Such solutions depend on a single scaling
variable $c = v/v_0(t)$ where $v_0(t)$ is the r.m.s. velocity or
instantaneous width of the distribution. This early evolution is
comparable to the rapid decay  of the distribution function to a
Maxwell-Boltzmann distribution in an spatially homogenous elastic
system. However, in systems of elastic particles similarity solutions
of the nonlinear homogeneous Boltzmann equation do {\it not} control
the long time behavior of $F(v,t)$ \cite{BKW}. The earlier studies
\cite{BDS,esipov,vNE98} of these problems were mainly focussing on
inelastic hard spheres, which is the proto-typical model for inelastic
gases and fluids, and on the extremely simplified inelastic BGK-models
with a single relaxation time \cite{BMD}.

More recently simplified stochastic models have been introduced
\cite{BN-K-00,Bobyl-00,Cercig-00} to tackle the nonlinear Boltzmann
equation, while keeping the essential physics of the inelastic
collisions. Unfortunately the microscopic dynamics of these stochastic
models is only defined for velocity variables, and the models can not
be studied in phase space using the $N-$particle methods of
statistical mechanics and molecular dynamics.

Nevertheless, the recent studies  of inelastic Maxwell models
\cite{Rome1,Rome2,ROME4,BN-K-02,BN-K-02a,BN-K-Springer,ME+RB,ME+RB-fest,ME+RB-rapid,EB-Springer,BC-WNtail,BCT02,GPV03,BGV03}
have greatly advanced our understanding of kinetic theory for
inelastic systems, as well as  the structure of the resulting velocity
distributions and the significance of scaling solutions, which are
exposing the generic features of relaxation both in homogeneous
cooling states, as well as in driven systems.

The physical importance of these scaling solutions has been
demonstrated by Baldassarri et al.~\cite{Rome1,Rome2,ROME4} with the
help of MC simulations of the nonlinear Boltzmann equation for one-
and two-dimensional systems of inelastic Maxwell particles. They found
that the solution, $F(v,t)$, after a short transient time,  could be
collapsed on a scaling form $v_0^{-d}(t) f(v/v_0(t))$ for large
classes of initial distributions $F(v,0)$ (e.g. uniform or Gaussian).
Moreover, in one dimension they found a simple exact scaling solution
of the Boltzmann equation, which has a heavily overpopulated algebraic
tail $\sim v^{-4}$ when compared to a Gaussian. In two dimensions they
have shown that the solutions of the initial value problem for regular
initial distributions (say, without tails)  also approach a scaling
form with over-populations in the form of algebraic tails, $f(c) \sim
c^{-d-a}$ with an exponent $a(\alpha)$ that depends on the degree of
inelasticity $\alpha$.

In the same period Krapivsky and Ben-Naim
\cite{BN-K-02,BN-K-02a,BN-K-Springer}, and independently the present
authors \cite{ME+RB,ME+RB-fest,EB-Springer} developed analytic methods
to determine the scaling solutions of the nonlinear Boltzmann equation
for freely evolving Maxwell models, and in particular to calculate the
exponent $a(\alpha)$ of the high energy tails $f(c) \sim 1/c^{a+d}$,
as a function of the coefficient of restitution $ \alpha$. Using the
methods of Ref. \cite{vNE98} the present authors have also extended
the above results to inelastic Maxwell models driven by different
modes of energy supply \cite{ME+RB-rapid}. Here the high energy tails
turned out to be stretched Gaussians, $f(c) \sim \exp[ -\beta c^b]$
with $0<b<2$. Very recently these studies were also generalized to
inelastic soft spheres \cite{EB-Springer,orsay} both for freely
evolving as well as for driven systems, where the over-populated high
energy tails turn out to be stretched Gaussians with $0<b<2$ as well.
This class of models covers both inelastic hard spheres and inelastic
Maxwell models as special cases. We will only briefly touch upon these
models in the concluding section of this article.

Subsequently there appeared also rigorous mathematical proofs of the
approach to these scaling forms for freely evolving inelastic Maxwell
gases \cite{BCT02}, for inelastic hard sphere gases driven by white
noise \cite{GPV03}, and for both types of systems driven by different
types of thermostats \cite{BGV03}.

This discovery stimulated a lot of theoretical and numerical research
in solutions of the Boltzmann equation, specially for large times and
for large velocities, as universal phenomena manifest themselves
mostly on such scales. Why were the first results all for Maxwell
models? Maxwell models \cite{BKW} derive their importance in kinetic
theory from the property that the collision frequency is independent
of the relative impact speed $g$, whereas the collision frequency in
general depends on the speed at impact. For instance, for hard spheres
the collision frequency is proportional  to $g$.

This property of Maxwell models simplifies the structure of the
nonlinear kinetic equation. For instance, the equations of motion for
the moments $\langle v^n \rangle$ can be solved sequentially as an
initial value problem; the eigenvalues and eigenfunctions of the
linearized collision operator can be calculated, and the
$(2d-1)-$dimensional nonlinear collision integral in the Boltzmann
equation can be reduced to a $(d-1)-$ dimensional one by means of
Fourier transformation \cite{Bobyl-00}. In subsequent sections we will
take advantage of these properties to determine similarity solutions
of the nonlinear Boltzmann equation for the $d-$dimensional IMM, and
the moment equations enable us to study the approach of initial value
solutions $F(v,t)$ to such similarity solutions.

The plan of the paper is as follows: In Sections 2 and 3 the nonlinear
Boltzmann equation for a one-dimensional gas of Maxwell particles is
solved  without and with energy input to obtain scaling solutions. The
high energy tails are respectively power laws or stretched Gaussian
tails. Section 4 is an intermezzo about an inelastic BGK-kinetic
equation without or with energy supply, for which the high energy
tails are analyzed.  After discussing in Section 5 the basics of the
nonlinear Boltzmann equation  for our two fundamental $d-$dimensional
inelastic models, i.e. inelastic Maxwell models (IMM) and inelastic
hard spheres (IHS), we repeat the above program  for $d-$dimensional
models in Section 6 for free cooling, and in Section 8 for driven
Maxwell models. In the intermediate Section 7 we study the approach of
the velocity distribution function  $F(v,t)$ towards the scaling
solution, and Section 9 gives some conclusions and perspectives.

\setcounter{section}{1}
\setcounter{subsection}{0}\setcounter{equation}{0}
\section{Freely cooling one-dimensional gases}
\subsection{Nonlinear Boltzmann equation}

Some of the basic features of inelastic systems can be discussed using
the Boltzmann equation for a simple one-dimensional model
\cite{BN-K-00} with inelastic interactions, possibly driven by
Gaussian white noise. Let us denote the isotropic velocity
distribution function at time $t$ by $F(v,t)=F(|v|,t)$. Its time
evolution is described by the nonlinear Boltzmann equation,
\be \Label{a1}
\frac{\partial}{\partial t}F(v,t)-D\frac{\partial^2}{\partial v^2}F(v,t) = I(v|F),
\ee
where the diffusion term, proportional to $\partial^2/\partial v^2$,
represents the heating effect of  Gaussian white noise with strength
$D$. The collision operator, $I(v|F)$, consists of two terms: the loss
term, that accounts for the so-called {\it direct} collisions of a
particle with velocity $v$ with any other particle, and the gain term
that accounts for the so-called {\it restituting} collision of two
particles with pre-collision velocities $v^{**}$ and $w^{**}$,
resulting in the post-collision velocities $(v,w)$:
\ba\Label{BE-1D}
I(v|F)&=&  \int dw\left[   \frac{1}{\alpha}
 F(v^{**},t)F(w^{**},t) - F(w,t) F(v,t)\right]\nonumber \\
&=&-F(v,t)+ \frac{1}{p}\int  duF(u,t)F \left(\frac{v-q u}{p},t\right),
\ea
where $p=1-q=(1+\alpha)/2$ and $\alpha$ is the coefficient of
restitution with $ 0 \leq \alpha \leq 1$. All velocity integrations in
\Ref{BE-1D} extend over the interval $(-\infty,+\infty)$. Because $F$
is normalized to unity, the loss term is simply equal to $-F(v,t)$.
Equation \Ref{BE-1D} is the nonlinear Boltzmann for the {\em Inelastic
Maxwell Model} (IMM) in one dimension, introduced by Ben-Naim and
Krapivsky \cite{BN-K-00}.

In a {\em direct collision}  two particles with velocities $v$ and $w$
collide, resulting in post-collision velocities, $v^*$ and $w^*$,
given by:
\ba\Label{coll-1d}
v^*= v(\alpha) \equiv \half(1-\alpha) v+\half(1+\alpha)w
&\mbox{\quad or\quad }& v^*=qv+pw  \nonumber \\
  w^*=w(\alpha) \equiv \half(1+\alpha)v+\half(1-\alpha) w
 &\mbox{or}&          w^*=pv+qw,
\ea
where total momentum is conserved. The {\em restituting collisions}
are events in which two particles with pre-collisional velocities
$v^{**}$ and $w^{**}$ collide such that one of the post-collisional
velocities is equal to $v$. The velocities are given by the inverse
transformation of \Ref{coll-1d}, i.e.
$v^{**}=v( 1/\alpha)$
and $w^{**}=w( 1/\alpha)$.

The velocity distribution function is normalized such that mass and
mean energy or granular temperature are,
\be\Label{Norm}
\int d\bv\, F(v,t) =  1; \qquad   \av{v^2}(t)=\int d\bv\, v^2 F(v,t)
\equiv  \half d v^2_0(t).
\ee
For later reference the normalization of the  mean kinetic energy is
written down for $d-$dimensional systems, where $v_0(t)$ is referred
as the r.m.s.\ velocity, and $T=v_0^2$ as the granular temperature.
The evolution equation of the energy $\langle v^2\rangle$ is obtained
by multiplying \Ref{a1} with $\int d\bv v^2$, which yields,
\be \Label{T-bal-1d}
\frac{d}{dt} v_0^2 = 4D -2pq v^2_0.
\ee
It describes the approach of $v_0^2(t)$  to a non-equilibrium steady
state (NESS) with r.m.s. velocity $v_0(\infty)= \sqrt{2D/pq}$. Here
the heating rate $D$ due to the random forces is balanced by the loss
rate, $\propto pq v^2_0$, caused by the inelastic collisions. The
energy balance equation \Ref{T-bal-1d} has an stable attractive fixed
point solution  $v_0(\infty)$ in the sense that any $v_0(t)$, with an
initial value different from $v_0(\infty)$, approaches this fixed
point at an exponential rate. This one-dimensional model Boltzmann
equation \Ref{a1} will be extended to general dimensionality $d$ in
later sections.

An inelastic fluid without energy input, in the so-called {\em
homogeneous cooling state}, will cool down due to the collisional
dissipation. In experimental studies of granular fluids energy has to
be supplied at a constant rate to keep the system in a non-equilibrium
steady state, while in analytic, numerical and simulation work freely
cooling systems can be studied directly. Without energy input the
velocity distribution $F (v, t)$ will approach a Dirac delta function
$ \delta(\bv) $ as $t\to\infty$, and all moments approach zero,
including the width $v_0 (t)$. However, an interesting structure is
revealed when velocities, $c = v/v_0 (t)$, are measured in units of
the instantaneous width $v_0 (t)$, and the long time limit is taken
while keeping $c$ constant, the so-called {\em scaling limit}.
{\em Scaling} or similarity solutions of the Boltzmann
equation are special solutions of a single scaling argument
$c=v/v_0(t)$, where the normalization of mass requires,
\be \Label{sim-sol}
F(v,t) = (v_0(t))^{-d} f (v/v_0 (t)) .
\ee
Consequently $f(c)$ satisfies the normalizations,
\be\Label{norm-scale}
\int d\bc f(c) = 1;\qquad \int d\bc \,c^2 f(c) =\half d.
\ee

To understand the physical processes involved we first discuss in a
qualitative way the relevant limiting cases. Without the heating term
$(D=0)$, equation (\ref{a1}) reduces to the freely cooling IMM. If one
takes in addition the elastic limit $q \to 0$, the collision laws
reduce in the {\it one-dimensional} case to $v^*=w,w^*=v$, i.e. an
exchange of particle labels, the collision term vanishes identically,
so that every $F(v,t) = F(v)$ is a solution, and  there is no
randomization or relaxation of the velocity distribution through
collisions, and the model becomes trivial at the Boltzmann level of
description. However, the distribution function in the presence of
{\it infinitesimal} dissipation $(\alpha \to 1)$ approaches a
Maxwellian. If we turn on the noise $(D \neq 0)$ at vanishing
dissipation $(q=0)$, the r.m.s.\ velocity in (\ref{T-bal-1d}),
$v^2_0(t)=v^2_0(0)+4Dt$, is increasing linearly with time. With
stochastic heating {\it and} dissipation (even in infinitesimal
amounts) the system reaches a NESS, through the balance of energy
input and dissipation.

\subsection{Scaling solutions}
After this qualitative introduction into the physical processes, we
will study scaling solutions \Ref{sim-sol} of the nonlinear Boltzmann
equation in a freely evolving system without energy input $(D=0)$. We
first take the Fourier transform of the Boltzmann equation \Ref{a1}.
Because of the convolution structure of the nonlinear collision term
\Ref{BE-1D} it reduces to
\be \Label{FT}
\frac{\partial}{\partial t}\Phi(k,t) =
-\Phi(k,t) + \Phi(kq,t)\Phi(kp,t).
\ee
 This equation possesses an interesting scaling or similarity
solution of the form, $\Phi(k, t) = \phi(kv_0(t))$, which is the
equivalent of \Ref{sim-sol} in Fourier representation. Substitution of
this ansatz into \Ref{FT} yields then,
\be \Label{FT-scale}
-pq k\phi'(k) =-\phi(k)+\phi(qk)\phi(pk),
\ee
where the energy balance equation \Ref{T-bal-1d} has been used to
eliminate $d v_0/dt $. By combining solutions of the form
$(1-sk)e^{sk}$ for positive and negative $k$, one can construct a
special solution of \Ref{FT-scale} as: $\phi(k)=(1+s|k|)\exp[-s|k|]$,
with an inverse Fourier transform \cite{Rome1} given by
\be    \Label{sol-1d}
f(c)=\frac{2}{\pi s}\frac{1}{[1+c^2/s^2]^2}.
\ee
By choosing $s=1/\sqrt{2}$, one obtains the scaling solution that
obeys the normalization $\av{c^2}=\half$, as imposed by  \Ref{Norm}.
The solution above has an algebraic form; it is even in the velocity
variable, with a power law tail $1/c^4$ at high energy, and moments of
order larger than 2 are all {\it divergent}. Notice that the solution
\Ref{sol-1d} does not depend on the inelasticity, which is an
exceptional property of the freely evolving  inelastic Maxwell model
in one dimension, as we shall see in later sections.

\subsection{Moment equations and approach to scaling}

An enormous simplification of both elastic and inelastic Maxwell
models, as compared to hard spheres or other interaction models, is
that the infinite set of moment equations can be solved sequentially
as an initial value problem. This enables us to investigate
analytically, at least to some extent, how the general solution $F(v,
t)$ of the complex nonlinear Boltzmann equation approaches the much
simpler scaling solution $f(c)$ of a single scaling variable.

   From general considerations we know already that the solution $F(v,
t)$ of the inelastic Boltzmann equation approaches a delta function.
So, all its moments must vanish in the long time limit, whereas the
moments $\av{c^n}$  of the scaling form are constant for $n = 0, 2$,
and divergent for $n > 2$  on account of \Ref{sol-1d}.

In our subsequent analysis it is convenient to define the standard
moments $m_n(t)$ and the rescaled moments $\mu_n(t)$ of the
distribution function as,
\be\Label{mom-def}
m_n(t)\equiv v_0^n \mu_n(t) =\frac{1}{n!} \int dv\, v^nF(v,t).
\ee
The  evolution equation of the moments can be obtained by multiplying
\Ref{a1} by $v^n$ and integrating over $v$, i.e.:
\be\Label{mom1}
\frac{d m_n}{dt} =- m_n +\frac{1}{p\,n!} \int\int dv  du\, v^n
F(u,t)F\left(\frac{v-q u}{p},t\right).
\ee
After a simple change of variables, the gain term in
 this equation transforms into,
\ba\Label{coll-mom}
\frac{1}{n!}\int\int du dw (pw+qu)^n F(u,t)F(w,t) = \sum_{l=0}^n p^l
q^{n-l} m_lm_{n-l},
\ea
where we have used Newton's binomial formula. We first observe that
particle conservation gives $m_0(t) = 1$. Combination of \Ref{mom1}
and \Ref{coll-mom}  yields the moment equations,
\be\Label{mom-final}
\frac{d m_n}{dt} +\lambda_n m_n = \sum_{l=2}^{n-2} p^l q^{n-l}
m_lm_{n-l},
\ee
where the eigenvalue $\lambda_n$ is given by,
\be\Label{eigenvalue}
\lambda_n=1-p^n-q^n .
\ee
Next we observe that for an isotropic $F(|v|,t)$ only the even moments
are non-vanishing. So, \Ref{mom-final}  only involves even values of
$l$ and $n$. This set of equations can be solved recursively for all
$n$. For $n = 2$ the nonlinear term on the right hand side vanishes,
and we find,
\be\Label{m2}
m_2(t) =m_2(0)e^{-\lambda_2 t}
\ee
with $\lambda_2= 2pq$. We note that $m_2(t)=v_0^2(t)/4$.  Similarly we
solve the equation for the fourth moment,
\be\Label{m4}
m_4(t)=\left[ m_4(0) +\half m_2^2(0)\right] \exp[-\lambda_4 t] - \half
m_2^2(0) \exp[-\lambda_2 t],
\ee
where the equality $\lambda_4 - 2\lambda_2 = - 2p^2 q^2$  has been
used. The coefficients in \Ref{m4} turn out to be independent of $p$
and $q$. Similarly we can show that the dominant long time behavior of
the higher even moments is given by $m_n(t) \sim \exp[ -\lambda_nt]$,
i.e. they decay asymptotically according to \Ref{mom-final} with the
nonlinear right hand side set equal to zero. This is the multi-scaling
behavior of the velocity moments found by Ben-Naim and Krapivsky in
\cite{BN-K-00,BN-K-Springer}. Consequently all  moments with $n > 0$
vanish as $t\to\infty$, consistent with the fact that the distribution
$F(v, t) \to\delta (v)$ in this limit.

The approach of $F(v,t)$  to a scaling form in $d-$dimensional
inelastic models was first formulated and conjectured  by the present
authors based on their analysis of the time evolution of the moments
(for inelastic hard spheres and inelastic Maxwell models see
\cite{ME+RB-fest}; extended to inelastic soft spheres
\cite{EB-Springer,orsay}), and was subsequently proven in a
mathematically rigorous fashion for inelastic Maxwell models and hard
spheres \cite{BCT02,GPV03,BGV03}. Rather than presenting the
mathematical proof it is of more interest from the physics point of
view to understand how the physically most important lower moments of
$F(v,t)$ approach their limiting form, and relate these limits to the
moments of the scaling solution \Ref{sol-1d}.

We first observe that the assumed large$-t$ behavior \Ref{sim-sol} of
$F(v,t)$ in combination with \Ref{mom-def} implies that $  \mu_n(t)
\to \mu_n$ for $n = 2, 4\dots$, where $\mu_n \equiv (1/n!) \int
dc\,c^n f(c)$. So, the moment equations for the rescaled moments
follow directly from \Ref{mom-final} and \Ref{mom-def} to be,
\be\Label{mom-scaled}
\frac{d\mu_n(t)}{dt} +\gamma_n \mu_n(t)= \sum_{l=2} ^{n-2} p^l q^{n-l}
\mu_l(t) \mu_{n-l}(t),
\ee
with eigenvalue
\be\Label{gamma-n}
\gamma_n=\lambda_n -\half n\lambda_2
\ee
on account of \Ref{mom-final} and \Ref{m2}. For $n=2$ one obtains that
$\mu_2 (t) = \mu_2 = \half \av{c^2} =\textstyle{\frac{1}{4}}$ is
constant for all times, in agreement with the corresponding moment of
the scaling form \Ref{sol-1d}. Next we consider the  fourth moment
$\mu_4(t)$, which  follows immediately from \Ref{m4} and \Ref{mom-def}
and reads,
\be\Label{reduced-m4}
\mu_4(t)=m_4(t)/v_0^4(t) = -\half \mu^2_2 + \exp(2p^2q^2 t)[\mu_4(0)
+\half \mu_2^2].
\ee
This solution is indeed {\it positive} and finite for all finite
positive times, and approaches $+\infty$ as $t$ becomes large. A
similar   result for the time dependence of  $\mu_n(t)$ is obtained
from \Ref{mom-scaled} for $n=6,8, \cdots$. This behavior is fully
consistent with the exact scaling solution \Ref{sol-1d}, of which all
even moments with $n>2$ are divergent.

Now, a curious result follows by considering the stationary solutions
of \Ref{mom-scaled} by setting $d\mu_n(t)/dt=0$. The equations then
reduce to a simple recursion relation,
\be \Label{mu-recurs}
\mu_n=\frac{1}{\gamma_n} \sum_{l=2} ^{n-2} p^l q^{n-l}\mu_l \mu_{n-l}
\ee
with initialization, $\mu_2 = \textstyle{\frac{1}{4}}$. This yields
for $n = 4$,
\be\Label{mu4}
\mu_4= \frac{p^2 q^2}{\gamma_4} \mu_2^2 =-\displaystyle{\frac{1}{32}},
\ee
where we have used the relation $\gamma_4=-2p^2q^2$, as follows from
\Ref{eigenvalue}. A {\em negative} moment of a physical (positive)
distribution is clearly {\em unphysical}. This unphysical behavior
continues for higher moments, where one finds in a similar manner that
the even moments approach a {\it finite} value, $\mu_{2n} \propto
(-)^{n + 1} {\cal C}_n$, with alternating signs!

What is happening here? Clearly, the set of solutions $\{\mu_{2n} \}$
of the limiting equation \Ref{mu-recurs} {\it differs} from the long
time limit $\mu_{2n}(\infty)$ of the set  of solutions $\mu_{2n}(t)$
of \Ref{mom-scaled}. The fixed point solution $\{\mu_n, n=2,4,\dots\}$
of \Ref{mu-recurs} does not represent a {\it stable attracting} fixed
point for physical solutions, but an unstable/repelling one. The
physical solutions $\mu_n(t)$ of \Ref{mom-scaled} move for large $t$
away from the unstable fixed point $\{\mu_n\}$ at an exponential rate,
given by $\exp(2p^2q^2t)$.

\setcounter{section}{2}
\setcounter{subsection}{0}\setcounter{equation}{0}
\section{Driven 1-D gases}

In this section we discuss the scaling form of the velocity
distribution in the non\-equilibrium steady state (NESS) for the
one-dimensional IMM for different modes of energy supply, and the
results are compared with those of the proto-typical dissipative fluid
of hard spheres in one dimension. In general, a NESS can be reached
when the energy loss through collisional dissipation is compensated by
energy supplied externally, as described in \Ref{T-bal-1d} for the
case of external Gaussian white noise. This stationary case is
described by the Boltzmann equation \Ref{a1} with $\partial_t
F(v,t)=0$. Moreover, at the end of this section all known results for
scaling forms in {\it driven} inelastic one-dimensional systems will
be summarized.

To obtain a description of the NESS, that is independent of the
details of the initial state, we rescale velocities, $
c=v/v_0(\infty)$, in terms of the r.m.s.~velocity $v_0(\infty)$, and
express $F(v,\infty)$ in terms of the scaling form $f(c)$ introduced
in \Ref{sim-sol}. The Boltzmann equation \Ref{a1} then takes the form,
\be \Label{b1}
-\frac{D}{v^2_0(\infty)} \frac{d^2f}{d c^2} = - \half pq \:d^2 \!
f(c)/d c^2 = I(c|f),
\ee
where $v_0(\infty)=\sqrt{2D/pq}$. The integral equation for the
characteristic function, $\phi(k) =\int dc\, e^{-ikc} f(c)$, follows by
taking the Fourier transform of the above equation,
\be \Label{b2}
(1+\half \ pqk^2) \phi(k) = \phi(pk)\phi(qk),
\ee
where the nonlinear collision term has been obtained as in \Ref{FT}.
An exact scaling solution of this equation in the form of an infinite
product has been obtained by Ben-Naim and Krapivsky
\cite{BN-K-00,BN-K-Springer}, and by Nienhuis and van der Hart
\cite{Nienhuis}. Here we construct the solution following Santos
and Ernst in \cite{SE-condmat}. We multiply \Ref{b2} on both sides
with $\phi_0(k) \equiv 1/(1+\half pq k^2)$, and write the equation as
an iteration scheme,
\be \Label{sol-iterat}
\phi_{n+1} =\phi_0(k)\phi_n(pk)\phi_n(qk).
\ee
The solution can be found iteratively  by starting from $\phi_0(k)$,
and inserting $\phi_n(k)$ on the right hand side to obtain
$\phi_{n+1}(k)$ on the left hand side. The result is the infinite
product,
\begin{equation} \Label{infty-prod}
\phi(k)=\prod_{m=0}^\infty\prod_{\ell=0}^m
\left(1+k^2/k_{m\ell}^2\right)^{-\nu_{ml}},
\end{equation}
where $\nu_{m\ell}\equiv \combin{m}{\ell}$ and $k_{m\ell}\equiv
ap^{-\ell}q^{-(m-\ell)}$ with $a \equiv \sqrt{2/pq}$. Thus $\phi(k)$
has infinitely many poles at $k=\pm i k_{m\ell}$ on the imaginary axis
with multiplicity $\nu_{m\ell}$ (for $\alpha \neq 0$). The velocity
distribution,
\begin{equation}\Label{c2}
f(c)=\frac{1}{2\pi}\int_{-\infty}^\infty dk\, e^{i kc}\phi(k),
\end{equation}
can then be obtained by contour integration. As $f(c)$ is an even
function, we only need to evaluate the integral in (\ref{c2}) for
$c>0$. The replacement $c\to|c|$ gives then the result for all $c$. By
closing the contour by a semi-circle through the infinite upper
half-plane and applying the residue theorem we obtain the exact
solution in the form of an infinite sum over poles.
This representation of $f(c)$ is very well suited to discuss the high
energy tail.

The dominant terms  for large $|c|$ correspond the poles closest to
the real axis, i.e. to the smallest values of $k_{m\ell}$. In case
$\alpha \neq 0$ the two smallest values are $k_{00}=a$ and
$k_{11}=a/p$. Consequently, the leading and first subleading term are
\begin{equation}\Label{c6}
f(c)\approx A_0e^{-a|c|}+A_1e^{-(a/p)|c|}+\cdots,
\end{equation}
where $a=\sqrt{2/pq}$ and the coefficients are found as
\ba \Label{c7}
A_0 &=&\frac{a}{2}\prod_{n=0}^\infty \exp\left[ \left(
\frac{1}{n+1}\right)\frac{p^{2(n+1)}+q^{2(n+1)}}{1-
p^{2(n+1)}-q^{2(n+1)}}\right] \nn
A_1 &=&(-)\frac{ap^3}{2(1-p^2)(p-q)}\nonumber\\
&&\times\prod_{n=0}^\infty \exp\left[ \left(
\frac{p^{-2(n+1)}}{n+1}\right)
\frac{\left(p^{2(n+1)}+q^{2(n+1)}\right)^2}{1-
p^{2(n+1)}-q^{2(n+1)}}\right].
\ea
In conclusion, the scaling form $f(c)$ in driven one-dimensional IMM
systems shows again an {\it overpopulated} high energy tail $\sim
\exp(-a|c|)$, when compared to a Gaussian. However this overpopulation
is very much smaller than in the freely cooling case, where $f(c) \sim
1/c^4$ for large $c$.

\begin{figure}
 $$\includegraphics[width=0.6 \columnwidth]{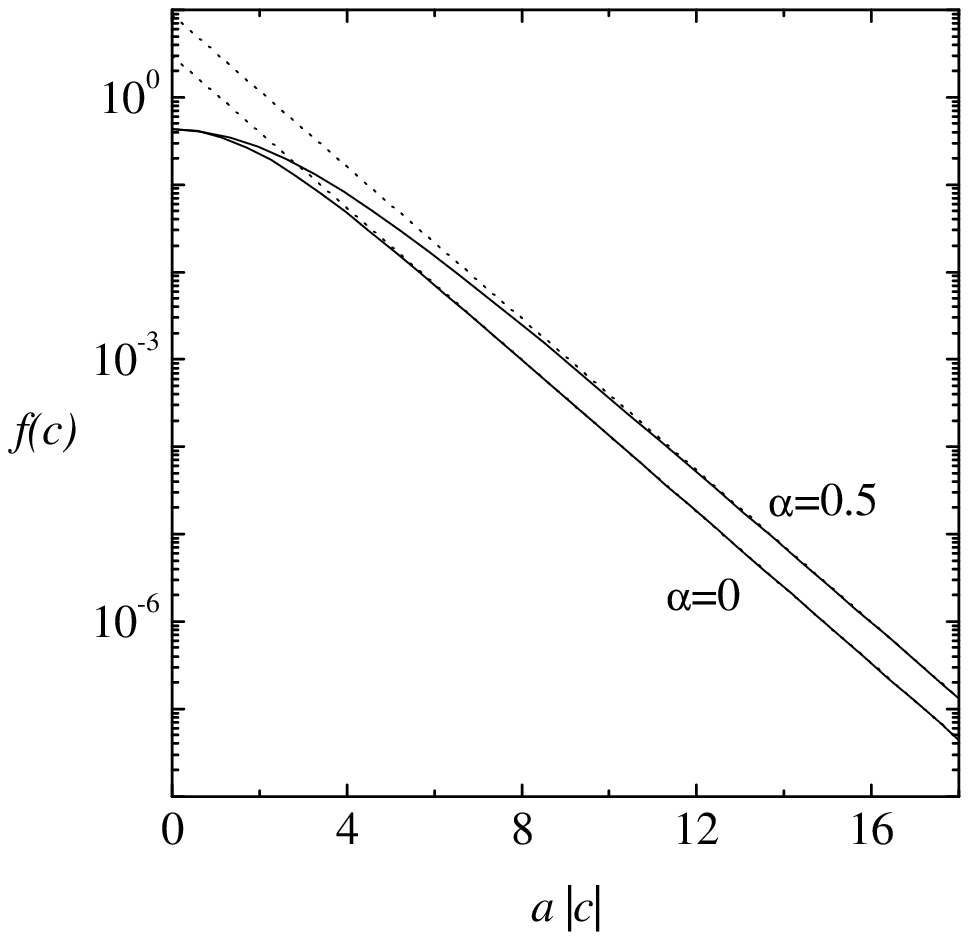}$$
\caption{\small Logarithmic plot of $f(c)$ versus $a|c|$ for $\alpha=0$ and
$\alpha=0.5$. The dotted lines are the asymptotic forms $f(c)\approx
A_0 e^{-a|c|}$ at $\alpha=0$ and $\alpha=0.5$, obtained from
(\ref{c7}). \Label{fig2}}
\end{figure}

Figure \ref{fig2} compares the asymptotic form $f(c)\approx A_0
e^{-a|c|}$ with the function $f(c)$ obtained by numerically inverting
$\phi(k)$ in \Ref{b2} for $\alpha=0$ and $\alpha=0.5$. Similar results
have been obtained by Nienhuis and van der Hart \cite{Nienhuis} and by
Antal et al. \cite{Droz}. We observe that the asymptotic behavior is
reached for $a|c| \largersim 4$ if $\alpha=0$ and for $a|c|\largersim
8$ if $\alpha=0.5$. As $a=\sqrt{2/pq}$ this corresponds to velocities
far above the r.m.s. velocity.

Interesting limiting behavior is also found in the {\it quasi-elastic}
limit. This limit is much more delicate and requires some care. If we
first take $\alpha\to 1$ at fixed $|c|$ and next $|c|\to\infty$
(denoted as order A in Table I), the high energy tail has a Maxwellian
form. On the other hand, if the limits are taken in the reverse order,
i.e.\ first $|c|\to\infty$ at fixed $\alpha<1$ and then $\alpha\to 1$
(denoted as order B in Table I), the asymptotic high energy tail is
exponential. The crossover between both limiting behaviors is roughly
described by the coupled limit $c \to \infty$ and $q \to 0$ with the
scaling variable $w = |c| \sqrt{q} =$ fixed   with   $q \equiv \half
(1-\alpha)\ll 1$, and occurs at $w \simeq 1$.   If $w<1$  the
distribution function is essentially a Maxwellian, while the true
exponential high energy tail is reached if $w \gg 1$.

The results for the scaling form in the quasi-elastic limit not only
depend sensitively on the order in which both limits are taken. They
also depend strongly on the collisional interaction, i.e. on the
energy dependence of the collision frequency, as well as on the mode
of energy supply to the system.

\begin{table*}[t]
\begin{tabular}{llcc}
State&System &Order A&
Order B\\
\hline\\[-4mm] Free cooling&Hard
spheres$^{1,\,2}$ & $\frac{1}{2}\left[\delta(c+c_0)
+\delta(c-c_0)\right]$&$e^{-a|c|}$\\
 &Maxwell model$^{3,\,3}$ &$c^{-4}$&$c^{-4}$\\
White noise&Hard spheres$^{1,\,4}$&$e^{-a|c|^3}$&$e^{-a|c|^{3/2}}$\\
&Maxwell model$^{5,\,6}$&$e^{-ac^2}$&$e^{-a|c|}$\\
Gravity thermostat&Hard
spheres$^{5,\,7}$&$\frac{1}{2}
\left[\delta(c+c_0)+\delta(c-c_0)\right]$& $e^{-ac^2}$ \\
&Maxwell model$^{5,\,8}$&$\frac{1}{2}\left[
\delta(c+c_0)+\delta(c-c_0)\right]$& $e^{-a|c|}$\\[0.3mm]\hline
\end{tabular}

 \vspace{5mm}
\caption{\Label{table1} \small Asymptotic behavior of the 1-D scaling
form $f(c)$ in the quasi-elastic limit for orders A and B. The
first/second footnote in the second column gives the reference where
the result for order A/B was obtained. References: $^1$ {Ref.\
\protect\cite{BBRTW}}, $^2$ {Ref.\ \protect\cite{vNE98}}, $^3$ {Refs.\
\protect\cite{Rome1,BN-K-00,ME+RB}}, $^4$ {Refs.\
\protect\cite{vNE98,BBRTW}}, $^5$ {Ref.\protect\cite{SE-condmat}},
$^6$ {Refs.\ \protect\cite{BBRTW,ME+RB-rapid}}, $^7$
{Ref.\protect\cite{MS00}}, $^8$ {Ref.\ \protect\cite{ME+RB-rapid}}. }
\end{table*}

To illustrate this we have collected in Table I what is known in the
literature for the different inelastic models in one dimension: (i)
hard spheres and (ii) Maxwell models, and for different modes of
energy supply: (i) no energy input or free cooling, (ii) energy input
or driving through Gaussian white noise, represented by the forcing
term $-D \partial^2 F(v,t)/ \partial v^2$ in  Boltzmann equation
\Ref{a1}, and (iii) energy input through a {\it negative} friction
force $\propto \zeta v/|v|$, acting in the direction of the particle's
velocity, but independent of its speed. This forcing term, referred to
as gravity thermostat in \cite{MS00}, can be represented by a forcing
term, $ \zeta (d/dv)(v/|v|) F(v,t)$, in the Boltzmann equation.  The
results for the gravity thermostat corresponding to order A
have been obtained by the same method as followed in Ref.
\cite{BBRTW}.

Inelastic gases in one dimension exhibit the remarkable property that
single peaked initial distributions develop into double peaked
solutions as the inelasticity decreases \cite{BBRTW,pulverenti}. It is
worthwhile to note that in the quasi-elastic limit a bimodal
distribution, $\frac{1}{2}\left[ \delta (c+c_0) + \delta (c-c_0)
\right] $ with $c_0= 1/\sqrt{2}$, is observed in inelastic hard sphere
systems both for free cooling and for driving through the gravity
thermostat. In inelastic Maxwell models however, this bimodal
distribution is only observed for the gravity thermostat.

It is  also important to note that in the normalization where
velocities are measured in units of the r.m.s.~velocity, the high
energy tail in the driven inelastic Maxwell model is only observable
for very large velocities, as illustrated in Figure \ref{fig2} for
strong $(\alpha \to 0)$ and intermediate $(\alpha =\half)$
inelasticity. In the quasi-elastic limit, where $(\alpha \to 1)$, the
tail is even pushed further out towards infinity.  This also explains
how to reconcile the paradoxical results of {\it exponential}
large-$c$ behavior with the very accurate representation of the
distribution function in the thermal range $(c \simeq 1)$, in the form
of a Maxwellian, multiplied by a polynomial expansion in Hermite or
Sonine polynomials with coefficients related to the cumulants (see
Refs. \cite{SE-condmat}). The validity of these polynomial expansions,
over a large range of inelasticities with $(0 \leq \alpha <1)$ has
been observed before in \cite{vNE98} for inelastic hard spheres and in
\cite{Cercig-00} for inelastic Maxwell models. Derivations of the
results, collected in Table I can be found in the original literature.

\setcounter{section}{3}\setcounter{subsection}{0}\setcounter{equation}{0}
\section{Inelastic BGK-Model}

A brief intermezzo on a very simple inelastic Bhatnagar-Gross-Krook
(BGK) model, introduced by Brey et al.~\cite{BMD}, is presented here
to show to what extent this model captures the correct physics as
described by the nonlinear Boltzmann equation for $d-$dimensional
inelastic Maxwell models.

A crude scenario for relaxation without energy input has to show that
the system cools down due to inelastic collisions, and that the
velocity distribution $F\left(v,t\right)$ approaches a Dirac delta
function $\delta \left({\bv}\right)$ as $t\to\infty$, while the width
or r.m.s. velocity $v_0\left(t\right)$ of this distribution, defined
as $\langle v^2 \rangle = \frac{1}{2} d v_0^2 $, is shrinking.
Moreover with a constant supply of energy, the system should reach a
non-equilibrium steady state (NESS). These features are implemented in
a simple  BGK-equation, i.e.
\begin{equation} \label{h1}
\partial_t F\left(v,t\right) -D\nabla_{\bv}^2
F\left(v,t\right)= -\omega_0
\left[F\left(v,t\right)-F_0\left(v,t\right)\right].
\end{equation}
Here $\omega_0$ is the mean collision frequency, which is chosen here
as  $\omega_0=1$ in order to model Maxwell models. The terms
$-\omega_0 F(v,t)$ and $ \omega_0 F_0(v,t)$ model respectively the
loss and (nonlinear) gain term. This kinetic equation describes the
relaxation of $F\left(v,t\right)$ towards a Maxwellian with a width
proportional to $\alpha v_0\left(t\right)$, defined by
\begin{equation} \label{h2}
F_0\left(v,t\right)= \left(\sqrt\pi \alpha v_0\right)^{-d}
\exp\left[-\left(v / \alpha v_0\right) ^2\right] \equiv \left(\alpha
v_0\right)^{-d} \phi\left(c/\alpha \right),
\end{equation}
where $c=v/v_0$. The energy balance equation follows from  \Ref{h1}
as,
\be \Label{h3}
dv_0^2/dt = 4D-(1-\alpha^2)v_0^2 \equiv 4D -2 \gamma  v_0^2,
\ee
where $(1-\alpha^2)$ measures the inelasticity. In the case of {\it
free cooling} ($D=0$) the solution is $v_0(t)= v_0(0) \exp[-\gamma
t]$. By inserting the scaling ansatz \Ref{sim-sol} into \Ref{h1} we
obtain in the case of free cooling an ordinary differential equation,
that can be solved exactly \cite{BMD,EB-Springer,orsay}. Its high
energy tail is,
\be \Label{h4}
f(c) \sim A/c^{a+d} \quad \mbox{ with} \quad a=1/\gamma =
2/(1-\alpha^2),
\ee
where an explicit expression for the amplitude $A$ has been
calculated.

As we shall see in Section 6, a similar heavily overpopulated tail,
$f\left(c\right)\sim 1/c^{d+a}$ for $d>1$, has also be found in freely
cooling $d-$dimensional Maxwell models with $\omega_0=1$, where the
exponent $a\left(\alpha\right)$ takes for small inelasticity the form
$a \simeq 1/\gamma_0 = 4d /\left(1-\alpha^2\right)$.

For the BGK-model {\it driven} by external white noise  we obtain a
NESS with a rescaled distribution function  $F\left(v,\infty\right) =
v_0^{-d}\left(\infty\right) f\left(v/v_0\left(\infty\right)\right)$
with standard width $\langle c^2\rangle =\textstyle{\frac{1}{2}} d$,
and the rescaled equation for $f(c)$ is obtained from \Ref{h1}. Its
asymptotic solution for $c \gg \alpha$ is then,
\begin{equation}
\Label{h11}
 f\left(c\right)  \sim \exp\left[-\beta c^b\right]
 =\exp\left[- 2 c/\sqrt{1-\alpha^2}\right].
\end{equation}

As we shall see in Section 8, a similar exponential high energy tail
is found in the white noise driven Maxwell models  .

\setcounter{section}{4}
\setcounter{subsection}{0}\setcounter{equation}{0}
\section{\mbox{\boldmath$d$}-Dimensional inelastic Maxwell gases}

In this section we consider the spatially homogeneous  Boltzmann
equation for inelastic Maxwell models in $d$ dimensions without energy
input. In most models of inelastic particles total momentum is
conserved in binary collisions, and the models qualify as inelastic
fluids. The details of the collision dynamics are defined in Figure 2.
The nonlinear Boltzmann equation for a $d$-dimensional IMM when driven
by Gaussian white noise can again be written in  the form,
\be \Label{IMM-BE}
\frac{\partial}{\partial t}F(v,t)-D \nabla^2_\bv F(v,t) = I(v|F),
\ee
where the collision term is,
\be \Label{IMM-coll}
I(v|F)= \int_n \int d\bw \left[\frac{1}{\alpha} F(v^{**},t)
F(w^{**},t) -F(v,t) F(w,t)\right]
\ee
Here $\int_n (\cdots) = (1/\Omega_d) \int d\bn(\cdots)$ is an angular
average over a full $d$-dimensional unit sphere with a surface area
$\Omega_d = 2 \pi^{d/2}/\Gamma(d/2)$. The factor $(1/\alpha)$ in the
gain term of  \Ref{IMM-coll} originates from the Jacobian, $d\bv^{**}
d\bw^{**} = (1/\alpha)d\bv d\bw $. The direct and restituting
collisions are given by:
\ba \Label{dyn}  &\bv^{*}= \bv -\half(1+\alpha) (\bg\cdot \bn) \bn; \quad
&\bw^{*} = \bw +\half(1+\alpha) (\bg\cdot \bn) \bn,\nn &\bv^{**}= \bv
-\half(1+\textstyle{\frac{1}{\alpha}}) (\bg\cdot \bn) \bn; \quad
&\bw^{**} = \bw + \half(1+\textstyle{\frac{1}{\alpha}}) (\bg\cdot \bn)
\bn.
\ea
In one dimension the  dyadic product $\bn \bn$ can be replaced by
unity, so that the equations above reduce to  \Ref{coll-1d}.

\begin{figure}[t]
$$\includegraphics[angle=270,width=.48 \columnwidth]{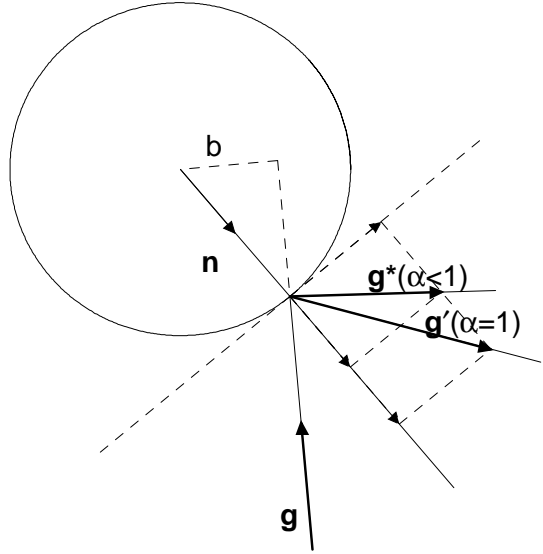}$$
\caption{\small Geometry of inelastic collisions, where $\bg =\bv -\bw$ is
the relative velocity, the unit vector $\bn$ specifies the point of
incidence on a unit sphere around the centre of force, $b=
|\hat{\bg} \times \bn| = |\sin \theta|$ is the impact parameter, and
$\theta$ the angle of incidence. In inelastic collisions the component
$g_\parallel = \bg \cdot \bn =g \cos \theta$ is reflected, i.e.
$g^*_\parallel = - \alpha g_\parallel$, and reduced in size by a
factor $\alpha = 1-2q = 2p-1$, where $\alpha$ is called the
coefficient of restitution $(0 \leq \alpha \leq 1)$, and $\alpha=1$
corresponds to the elastic case. If the total energy in a collision is
$E= \half (v^2 +w^2)$, then the energy loss in an inelastic collision
is  $\Delta E = - \fourth (1-\alpha^2) g_\parallel^2 = -pq
g_\parallel^2$. }
\end{figure}

To point out the differences between the Boltzmann equation for
inelastic Max\-well particles and the one for the prototypical inelastic
hard spheres we also quote  the Boltzmann collision term for inelastic
hard spheres, i.e.
\be\Label{IHS-BE}
I(v|F) = \int_n \int d\bw |{{\bg}} \cdot \bn |
\left[\frac{1}{\alpha^{2}} F(v^{**},t) F(w^{**},t) -F(v,t)
F(w,t)\right].
\ee
Here the collision term contains an extra factor $|\bg \cdot \bn|$,
and the gain term an extra factor $1/\alpha$ as compared to
\Ref{IMM-coll}. The velocity distributions in these inelastic models
with or without energy input were recently studied by many authors; in
particular the inelastic hard sphere gas in
\cite{williams+mackint,esipov,vNE98,Brey1,MS00,BBRTW,ME+RB-rapid} and
the inelastic Maxwell models in
\cite{BN-K-00,Bobyl-00,Cercig-00,Rome1,BN-K-02,ME+RB,ME+RB-fest,Nienhuis,Droz,SE-condmat}.

We return to the IMM and observe that in any inelastic collision an
amount of energy $\fourth(1-\alpha^2) g_\parallel^2 =pq g_\parallel^2$
is lost. Consequently the average kinetic energy or granular
temperature $v_0^2$  keeps decreasing at a rate proportional to the
inelasticity $\fourth(1-\alpha^2)=pq$. The balance equation can be
derived in a similar manner as \Ref{T-bal-1d}, and reads for general
dimensionality,
\be \Label{T-bal-dd}
\frac{d}{dt} v_0^2 = 4D-(2pq/d) v^2_0.
\ee
We will discuss {\it freely cooling} systems $(D=0)$ without energy
input, as well as {\it driven} systems (see Section 7), which can
reach a non-equilibrium steady state. In the case of free cooling the
solution of the Boltzmann equation does not reach thermal equilibrium,
described by a Maxwellian, but is approaching a Dirac delta function
$\delta^{(d)}(\bv)$ for large times. However, the arguments given in
Section 2  suggest again that $F(v,t)$ approaches a simple scaling
solution of the form $f(c)$, as defined in \Ref{sim-sol} after
rescaling the velocities as $c=v/v_0(\infty)$, with normalizations
chosen as in \Ref{norm-scale}.

In the case of free cooling the mean square velocity keeps decreasing
at an exponential rate, $v_0(t) =v_0(0) \exp[ -pqt/d]$, but the
distribution function rapidly reaches a (time independent) scaling
form $f(c)$, which is determined by the nonlinear integral equation,
\be \Label{FC-scale-BE}
\frac{pq}{d} \left(d+ c\frac{d}{dc}\right)f = \frac{pq}{d}  \bnabla_c \cdot
(\bc f) =I(c|f),
\ee
as can be derived by substituting \Ref{sim-sol} in \Ref{IMM-BE} and
rescaling  velocities.

For freely evolving and driven inelastic hard sphere fluids the
scaling solutions $f(c)$ have been extensively discussed both in the
bulk of the thermal distribution, as well as in the high energy tails
\cite{Brey1,vNE98,EB-Springer,orsay}.

\setcounter{section}{5}
\setcounter{subsection}{0}\setcounter{equation}{0}
\section{Scaling in \mbox{\boldmath{$d$}}-dimensional  free cooling}
\subsection{Fourier transform method}

To determine scaling solutions for free cooling $(D=0)$, we first
consider the Fourier transform of the distribution function,
$\Phi(k,t) = \langle \exp[-i\bk \cdot {\bv}]\rangle$, which is the
characteristic or generating function of the velocity moments $\langle
v^n \rangle$, and derive its equation of motion. Because
$F\left(v,t\right)$ is isotropic, $\Phi(k,t)$ is isotropic as well.

As an auxiliary step we first Fourier transform the gain term in
\Ref{IMM-coll}, i.e.
\ba
\Label{fourier-gain} &\int d\bv \exp[-i \bk \cdot \bv]
\:I_{\mbox{gain}}(v|F)=& \nn &\int_{n} \int d\bv d{\bw} \exp[-i \bk
\cdot \bv^{*}] F(v,t)F(w,t)
  = \int_{n} \Phi(k \eta_+,t)\Phi(k \eta_-,t).&
\ea
The transformation needed to obtain the first equality follows by
changing the integration variables $(\bv,\bw)\to(\bv^{**},\bw^{**})$
and using the relation $d\bv d\bw=\alpha d\bv^{**}d\bw ^{**}$. Then we
use \Ref{dyn} to write the exponent as $\bk\cdot \bv^* = \bk_- \cdot
\bv + \bk_+ \cdot \bw$, where
\ba
\Label{eta-pm} \bk_{+} &=  p (\bk \cdot \bn) \bn  \qquad | \bk_+| =k p
|(\hat{\bk}\cdot \bn)| = k \eta_+(\bn)
\nn \bk_{-}&  =\bk - \bk_{+} \qquad
|\bk_-| =k\sqrt{1-z(\hat{\bk}
 \cdot \bn)^2} = k\eta_-(\bn),
\ea
with $p=1-q=\textstyle{\frac{1}{2}} (1+\alpha)$ and $z=1-q^2$. In one
dimension $\eta_+(\bn) =p $ and $\eta_-(\bn) =q $, and $\int_{n} $ can
be replaced by unity. The Fourier transform of \Ref{IMM-coll} then
becomes the Boltzmann equation for the characteristic function,
\begin{equation} \Label{FT-BE}
\partial_t \Phi(k,t)= - \Phi(k,t) +
\int_{n} \Phi(k\eta_+(\bn),t) \Phi(k\eta_{-}(\bn),t),
\end{equation}
where $\Phi(0,t)=1$ because of the normalization of the distribution
function, and the collision operator is a $(d-1)-$dimensional
integral. In one-dimension this equation simplifies to \Ref{FT}.

Next we consider the moment equations. Because $F(v,t)$ is isotropic,
only its even moments are non-vanishing. If we assume that all moments
$\langle v^n \rangle$ are {\it finite}, then the moment expansion of
the characteristic function takes the form,
\begin{equation} \Label{mom-exp}
\Phi(k,t)= {\sum_{n}}^\prime \frac{(-ik)^n}{n!} \langle
(\hat{\bk}\cdot\bv)^n\rangle = {\sum_{n}}^\prime (-ik)^n m_n(t) ,
\end{equation}
and $\Phi(k,t)$  is a regular function of ${\bk}$ at the origin. In
the equation above the prime indicates that $n$ is even, and the
moments $m_n(t)$ are defined as,
\begin{equation} \Label{mom-n}
m_n (t) = \beta_n \langle v^n \rangle /n!,
\end{equation}
where $\beta_s = \int_{n} (\hat{\bk}\cdot\hat{\bf a})^s $ for real $s$
is an angular integral over $\bn$, given by:
\be \Label{beta-s}
\beta_{s} =
\frac{\int_0^{\pi/2} d\theta (\sin \theta)^{d-2} (\cos\theta)^{s} }
{\int_0^{\pi/2} d\theta (\sin\theta)^{d-2}} = \frac{
\Gamma(\frac{s+1}{2}) \Gamma(\frac{d}{2})}{
\Gamma(\frac{s+d}{2})\Gamma(\frac{1}{2})}.
\ee
Moreover, the normalizations \Ref{Norm} give $m_0(t)=1$ and $ m_2(t)=
\textstyle{\frac{1}{2}} \beta_2 \langle v^2\rangle = \fourth
v_0^2(t)$. By inserting the moment expansion \Ref{mom-exp} in the
Fourier transformed Boltzmann equation \Ref{FT-BE}, we obtain the
equations of motion for the coupled set of moment equations by
equating the coefficients of equal powers of $k$. The result is,
\begin{equation} \Label{mom-eq}
\dot{m}_n +\lambda_n m_n = \sum_{l=2}^{n-2} h\left(l,n-l\right)\: m_l\: m_{n-l}
\quad \left(n>2\right)
\end{equation}
with coefficients,
\ba
\Label{eigenval}& h(l,s) = \int_{n} \eta^l_+(\bn)\eta^s_-(\bn)&\nn
&\lambda_s = 1-h(s,0)-h(0,s)= \int_{n}
[1-\eta^s_+(\bn)-\eta^s_-(\bn)],&
\ea
where all labels $\{n,l,s\}$ take {\em even} values only. For later
use we calculate $ \lambda_2$ explicitly with the help of \Ref{eta-pm}
with the result  $\lambda_2 = 2pq/d $.

To obtain a scaling solution of \Ref{FT-BE}  we set $\Phi (k,t) =
\phi(kv_0(t))$ where the r.m.s.~velocity is obtained from
\Ref{T-bal-dd} with $D=0$, and reads  $v_0(t)= v_0(0) \exp (-pq t/d)=
v_0(0) \exp (-\half  t\lambda_2)$. This gives the {\it integral}
equation for the scaling form $\phi\left(k\right)$,
\begin{equation} \Label{phi-scale-eq}
-\half \lambda_2 k \frac{d}{dk} \phi\left(k\right) +\phi\left(k\right)
= \int_{n} \phi\left(k\eta_+\right)\phi\left(k\eta_-\right),
\end{equation}
which reduces for $d=1$  to,
\begin{equation} \Label{phi-scale-eq-1d}
-\half \lambda_2 k \frac{d}{dk} \phi(k) +\phi(k) = \phi(pk)\phi(qk).
\end{equation}
Here $\phi (k)$ is the generating function for the moments $\mu_n$ of
the scaling form $f(c)$, i.e.
\ba\Label{phi-k-exp}
\phi(k)& = & {\sum_n}^\prime \frac{(-ik)^n}{n!}
\beta_n \langle c^n \rangle \equiv
{\sum_n}^\prime (-ik)^n \mu_n \nonumber \\
&\simeq& 1 -\textstyle{\frac{1}{4}} k^2 + k^4
\mu_4 - k^6\mu_6 +\cdots,
\ea
where $n$ is even, $\mu_0=1$ and $ \mu_2 = \textstyle{\frac{1}{2}}
\beta_2 \langle c^2 \rangle = 1/4$ on account of the normalizations
\Ref{norm-scale} and $\beta_2 =1/d$ as given in \Ref{beta-s}.

\subsection{Small-$k$ singularity of the characteristic function}

In the previous section we have obtained the Boltzmann equation
\Ref{FT-BE} for the characteristic function, the moment equations
\Ref{mom-eq}, and the integral equation \Ref{phi-scale-eq} for the
scaling form $\phi (k)$. These equations provide the starting point
for explaining the MD-simulations of Baldassarri et al. for the freely
cooling IMM in two dimensions, i.e. data collapse after a short
transient time on a scaling form $f(c)$ with a power law tail. In the
present section we derive the solution $f(c)$ with a power law tail,
and in Section 7 we will study the approach to this scaling form.

The strategy to determine analytically  a possible solution with a
power law tail is by assuming that such solutions exist, then
inserting the ansatz $f(c) \sim 1/c^{a+d}$ into the scaling equation
\Ref{phi-scale-eq}, and determining the exponent $a$ such that the
ansatz is indeed an asymptotic solution.  We proceed as follows.
Suppose that $f(c) \sim 1/c^{a+d}$, then the moments $\mu_n$ of the
scaling form $f(c)$ are convergent for $n<a$ and are divergent for $n>
a$. As we are interested in physical solutions which can be {\em
normalized}, and have a {\em finite energy}, a possible value of the
power law exponent must obey $a>2$.

The characteristic function is in fact a very suitable tool for
investigating this problem. Suppose the moment $\mu_n$ with $n>a$
diverges, then the $n$-th derivative of the corresponding generating
function also diverges at $k=0$, i.e. ${\phi}\left(k\right)$ has a
singularity at $k=0$. Then a simple rescaling argument of the inverse
Fourier transform shows that $\phi\left(k\right)$ has a dominant
small-$k$ singularity of the form  ${\phi}\left(k\right) \sim
\left|k\right|^a$, where $a \neq even$. On the other hand,  when all
moments are finite, the characteristic function $\phi(k) $ is {\it
regular} at the origin, i.e. can be expanded in powers of $k^2$.

We first illustrate our analysis for the one-dimensional case. As the
requirement of finite total energy imposes the lower bound $a>2$ on
the exponent, we make the ansatz, consistent with \Ref{phi-k-exp},
that the dominant small-$k$ singularity has the form,
\begin{equation}\Label{ansatz}
\phi\left(k\right) = 1- \textstyle{\frac{1}{4}} k^2 + A\left|k\right|^a,
\end{equation}
insert this in \Ref{phi-scale-eq-1d}, and equate the coefficients of
equal powers of $k$. This yields,
\begin{equation} \Label{transc-1d-eq}
\half a \lambda_2= \lambda_a \quad \mbox{or} \quad a pq = 1-p^a-q^a.
\end{equation}
The equation has two roots, $a=2,3$, of which $a=3$ is the one larger
than 2. Here $A$ is left undetermined. Consequently the
one-dimensional scaling solution has a power law tail, ${f}(c) \sim
1/c^4$, in agreement with the exact solution \Ref{sol-1d}.

For general dimension we proceed in the same way as in the
one-dimensional case. We insert the ansatz \Ref{ansatz} into
\Ref{phi-scale-eq}, and equate the coefficients of equal powers of
$k$. This yields for the coefficient of $k^2$ the identity $2pq/d
=\lambda_2$, and for the coefficient of $k^a$ the transcendental
equation,
\begin{equation} \Label{trans-eq}
  \half a \lambda_2 =\int_{n} \left[1-\eta^a_+ -\eta^a_-\right] =\lambda_a.
\end{equation}
The equation above obviously has the solution $a=2$. We are however
interested in the solution with $a>2$. In the elastic limit ($\alpha
\to 1$) the solution is simple. There $ q \to 0$ and $a$ diverges. The
contributions of $\eta_\pm^a$ on the right hand side vanish because
$\eta_\pm <1$, and the exponent has the form,
\begin{equation} \Label{a-infty}
a \simeq \frac{d}{pq} = \frac{4d}{1-\alpha^2}.
\end{equation}
This result has qualitatively the same shape as the numerical solution
of \Ref{trans-eq}, shown in Figure 3 for $d=2$. Moreover the
simplified BGK-model of Section 4 predicts the same qualitative
behavior for the exponent of the power law tail. For general values of
$\alpha$ one needs to evaluate the integrals $h(a,0)$ and $h(0,a)$,
defined in \Ref{eigenval}.

\begin{figure}[ht]
\Label{aalphaexp}
 $$\includegraphics[angle=0,width=.65 \columnwidth]{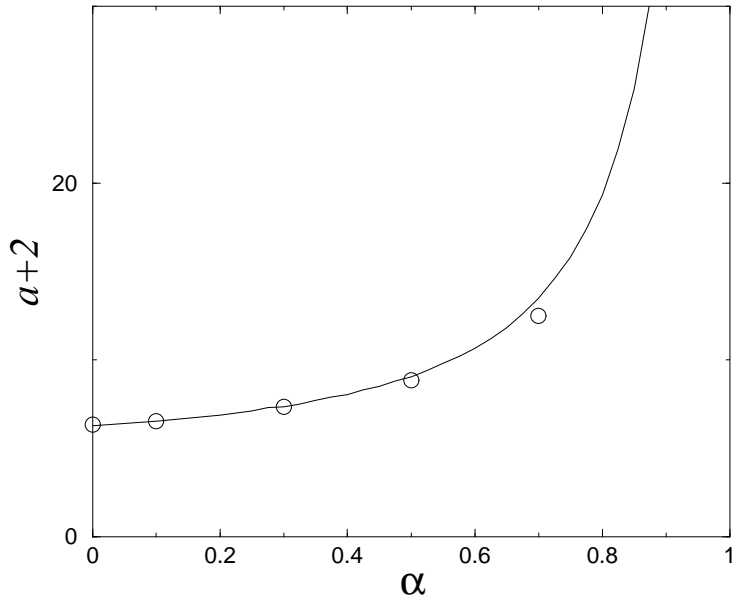}$$
\caption{\small Exponent $ a(\alpha)$ (solid line) as obtained by
numerical solution of \protect{\Ref{trans-eq}}. The open circles
represent the exponents, measured from the velocity distribution
functions obtained from MC simulations, and shown in Fig. 5. Figure
from Ref.\protect{\cite{orsay}}. }
\end{figure}

Here we only quote the results, and refer for technical details to the
literature \cite{ME+RB-fest,BN-K-02}, i.e.
\begin{equation} \Label{coeff-trans}
h\left(a,0\right) = p^a \beta_a; \qquad h\left(0,a\right) = \:
 _2F_1\textstyle{\left(-\frac{a}{2},\frac{1}{2}; \frac{d}{2};z\right)},
\end{equation}
where $\beta_a$ is given in \Ref{beta-s}, and $_2F_1$ is the
hypergeometric function with $z=1-q^2$.  One can conveniently use an
integral representation of $_2F_1$ to solve this transcendental
equation numerically.  We illustrate the solution method of
\Ref{trans-eq} with the graphical construction in Figure 4, where we
look for intersections of the line $y = \half s \lambda_2= \gamma_0 s
$ with the curve $y=\lambda_s$ for different values of $\alpha$.

\begin{figure}[ht]
\label{figaalpha}
 $$\includegraphics[width=.65 \columnwidth]{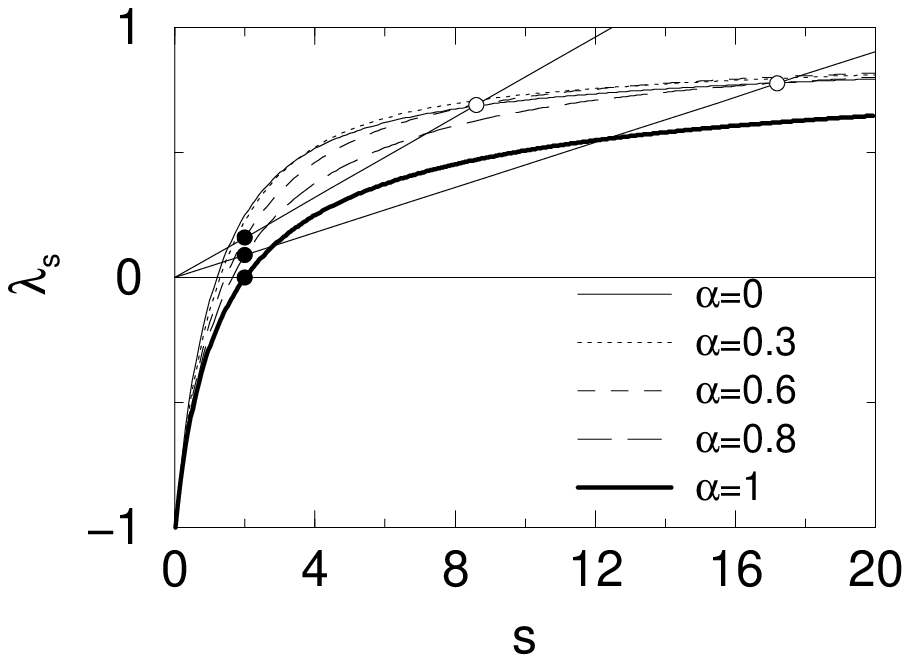}$$
\caption{\small Graphical solution of (\protect{\ref{trans-eq}}) for
different values of the paramenter $\alpha$. The eigenvalue
$\lambda_s$ is a concave function of $s$, plotted for different values
of the restitution coefficient $\alpha$ for the 2-D inelastic Maxwell
model. The line $y=s\gamma_0 =\half s \lambda_2$ is plotted for
$\alpha=0.6,0.8$ and $\alpha =1$(top to bottom). The intersections
with $\lambda_s$ determine the points $s_0$ (filled circles) and $s_1$
(open circles). Here $s_1=a$ determines the exponent of the power law
tail. For the elastic case ($\alpha=1, q=0$)
there is only one intersection point.}
\end{figure}

The relevant properties  of $\lambda_s $ are: (i)  $\lim_{s \to 0}
\lambda_s =-1$ whereas $\lambda_0=0$ because of particle conservation;
(ii) $\lambda_s$ is a concave function, monotonically increasing with
$s$, and {\em (iii)} all eigenvalues for positive {\em integers} $n$
are positive (see Figure 4). As can be seen from the graphical
construction, the transcendental equation \Ref{trans-eq} has two
solutions, the trivial one $(s_0=2)$ and the solution $s_1=a$ with
$a>2$. The numerical solutions $s_1(\alpha)$ for $d=2$ are shown in
Figure 3 as a function of $\alpha$, and the $\alpha$-dependence of the
root $a(\alpha)$ can be understood from the graphical construction. In
the elastic limit as $\alpha \uparrow 1$ the eigenvalue
$\lambda_2(\alpha)\to 0$ because of energy conservation. In that limit
the transcendental equation \Ref{trans-eq} no longer has a solution
with $a>2$, and $a(\alpha)\to\infty$ according to \Ref{a-infty}, as it
should be. This is consistent with a Maxwellian tail distribution in
the elastic case. Krapivsky and Ben-Naim \cite{BN-K-02} have also
solved the transcendental equation asymptotically for large $d \:(d
\gg 1)$, which gives qualitatively the same results as those shown in
Figure 3 for two dimensions.

These results establish the existence of scaling solutions
$f\left(c\right) \sim 1/c^{d+a}$ with algebraic tails, where the
exponent $a$ is the solution of the transcendental equation
\Ref{trans-eq}. Using a somewhat different analysis Krapivsky and
Ben-Naim \cite{BN-K-02} obtained the same results for the algebraic
tails in freely cooling Maxwell models.

The previous results have been confirmed in a quantitative manner by
means of MC simulations in \cite{ROME4,orsay} for different values of
$\alpha$. The algebraic tails of $f(c)$ are shown in Fig.~5, and the
exponents $a$, measured from  the MC data in Figure 5 are plotted in
Fig.~3. Both graphs  show excellent agreement with the analytic
results, derived here.

\begin{figure}[ht]
\Label{fc}
 $$\includegraphics[angle=0,width=0.68 \columnwidth]{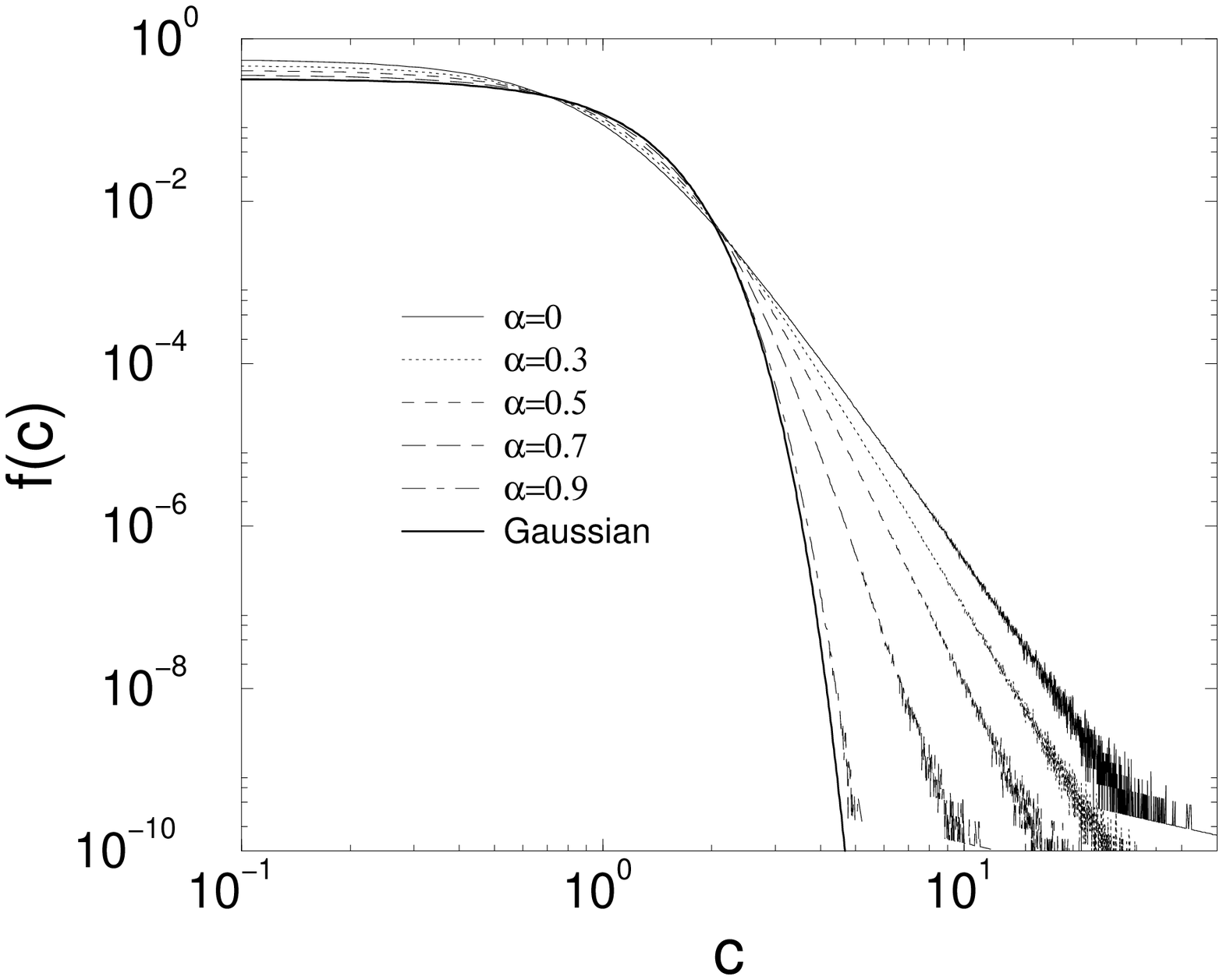}$$
\caption{\small Simulation of the velocity distribution function by
the MC method, showing power law tails. Figure from Ref.
\protect{\cite{orsay}}. }
\end{figure}

\setcounter{section}{6}
\setcounter{subsection}{0}\setcounter{equation}{0}
\section{Moment equations and approach to scaling}
\subsection{ Moments of velocity distributions}

In this section we study the effects on the moments of power law tails
in the scaling form, and we analyze in what sense the {\it even}
moments $m_n(t) =
\beta_n \langle v^n \rangle / n!$  at large times are related to the
moments $\mu_n = \beta_n \langle c^n \rangle / n! $ of the scaling
form $f(c) \sim 1/c^{a+d}$, which are {\it divergent } for $n>a$ and
remain {\it finite} for $n<a$.

First consider the moment equations \Ref{mom-eq} where $m_0(t)=1$ and
$ m_2(t) = m_2(0) \exp[- \lambda_2 t] = \fourth v^2_0(t)$. Similarly
one shows \cite{ME+RB-fest,BN-K-02} that $m_n(t) \sim \exp[-\lambda_n
t]$ for large $t$. Consequently all moments  with $n>0$ vanish as $t
\to \infty$, consistent with the fact $F(v,t) \to \delta^{(d)}(\bv)$
in this limit.

The rescaled moments $\mu_n(t) \equiv m_n(t)/v^n_0(t)$ show a more
interesting behavior. We analyze how they approach  their limiting
form $\mu_n(\infty)$. Inserting this definition into the moment
equations \Ref{mom-eq} and using $v_0(t) = v_0(0) \exp[-\half
\lambda_2 t]$ we find for the rescaled even moments with $n>0$,
\ba \Label{A7}
\dot\mu_n(t) +\gamma_n \mu_n(t) &=& \sum_{l=2}^{n-2} h(l,n-l)
\mu_l(t)\mu_{n-l}(t)
\nn \gamma_n  &=& \lambda_n -\half n\lambda_2 .
\ea

\begin{figure}[ht]
\Label{fig-moments}
 $$\includegraphics[width=0.65 \columnwidth,angle=270]{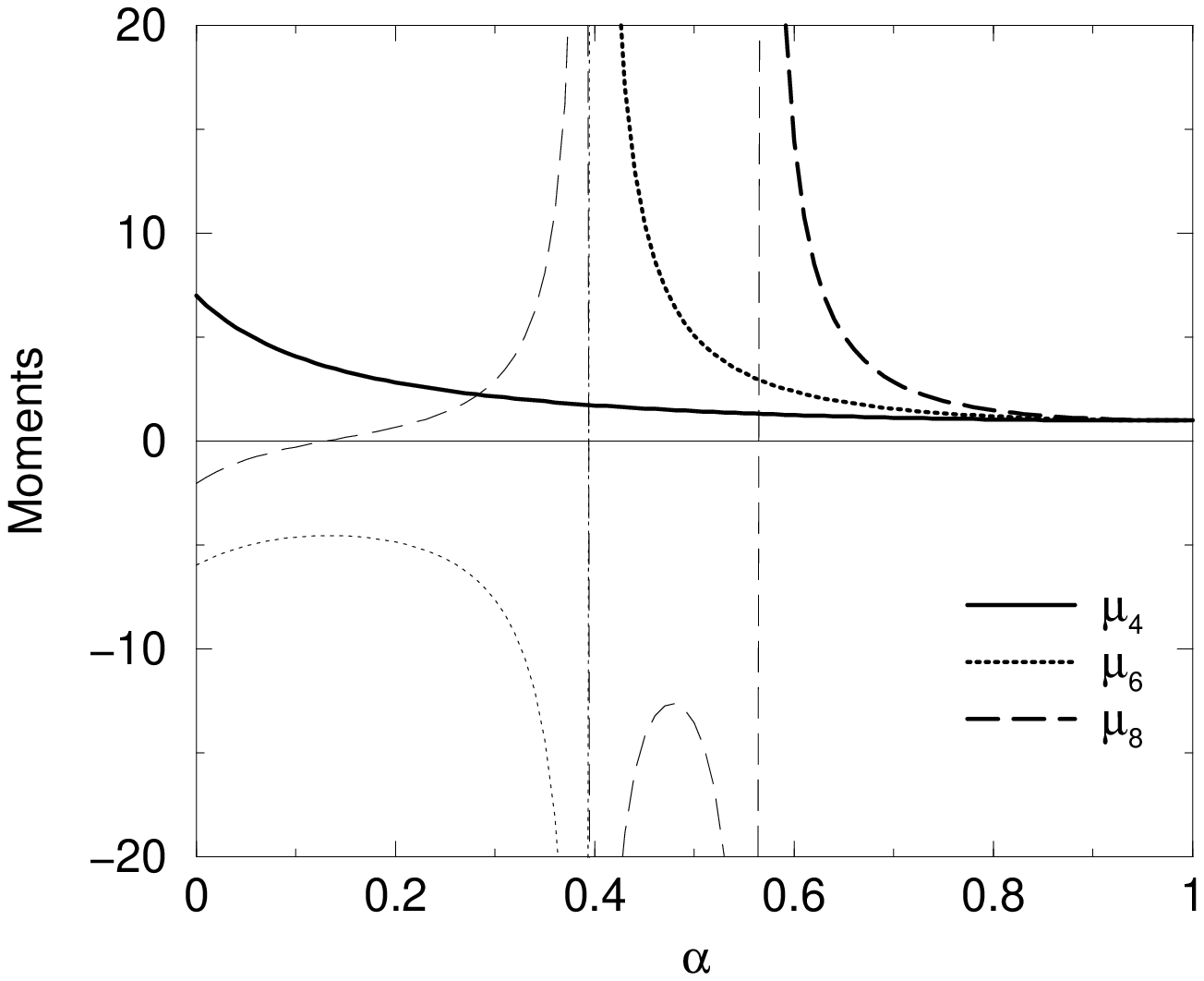}$$
\caption{\small  Moments $\mu_4$, $\mu_6$ and $\mu_8$ as a function of
the coefficient of restitution $\alpha$. Starting at $\alpha=1$  the
moment $\mu_n$ increases monotonically as $\alpha$ decreases following
the physical branch (thick line) untill $\alpha$ reaches a zero of
$\gamma_s$, where $\mu_n$ diverges towards $+\infty$. The recursion
relation \Ref{A5} has a second set of solutions $\{\mu_n\}$ that
become negative for small $\alpha$, indicating unphysical solutions.}
\end{figure}

The infinite set of moment equations \Ref{A7} for $\mu_n(t)$ can be
solved sequentially for all $n$ as an initial value problem. To
explain what is happening, it is instructive to consider again the
graphical solution of the equation, $\gamma_s=\lambda_s- \half
s\lambda_2=0$ for different values of the inelasticity $q$ or
$\alpha$, as illustrated by the intersections $\{s_0,s_1\}$ of the
curve $y=\lambda_s$ and the line $y=\half s\lambda_2$, where $s_0$ and
$s_1$ are denoted respectively by filled ($\bullet$) and open circles
($\circ$). These circles divide the spectrum into a {\em stable}
branch ($s_0 <s<s_1$) and two {\em unstable} branches ($s<s_0$) and
($s>s_1$). The moments $\mu_s(t)$ with $s=n
>a$ are on an unstable branch $(\gamma_s <0)$ and will grow for
large $t$ at an exponential rate, $ \mu_n(t) \simeq \mu_n(0)
\exp[|\gamma_n|t]$, as can be shown by complete induction from
\Ref{A7} starting at $n=[a] +1$. They remain positive and finite for
finite time $t$, but approach $+ \infty$ as $t\to\infty$, in agreement
with the predictions of the self consistent method of Section 6. The
moments  $\mu_n$ with $n=2,4, \cdots, [a]$ with $n$ on the stable
branch are globally stable and approach for $t\to\infty$ the limiting
value $\mu_n(\infty)=\mu_n$,  which are the finite positive moments of
the scaling form \Ref{A4}, plotted in Fig. 6. In summary, $\mu_n(t)
\to \infty$ if $n>a$, and $\mu_n(t)$ approach $\mu_n(\infty) = \mu_n$
for $n<a$, in agreement with the predictions of the power law tails in
Section 6.2.

The behavior of the moments described above is considered as a weak
form of convergence or approach of the distribution function $F(v,t)$
to the scaling form $f(c)$ for $t \to \infty$. The physically most
relevant distribution functions are those with {\em regular} initial
conditions, i.e. all moments $m_n(0)= v_0^n(0) \mu_n(0) < \infty$.

\subsection{Moments of scaling forms}
Next we consider the moments $\mu_n$ generated by the scaling form
$\phi(k)$ in \Ref{ansatz}, corresponding to $f(c) \sim 1/c^{a+d}$,
where $a>2$ and not equal to an even integer.   This implies that the
$n-$th order derivatives of $\phi(k)$ at $k=0$, and equivalently all
moments $\mu_n$, are {\em finite} if $n \leq [a] < a$, and all those
with $n>a$ are {\em divergent}. Here $[a]$ is the largest integer less
than $a$  where $[a]$ may be an  even or odd integer. Hence, the
{\em small}-$k$ behavior of $\phi(k)$ can be represented  as ,
\be \Label{A4}
\phi(k) =   {{\sum_{n=0}^{[a]}}}\,\!^\prime (-k)^n \frac{\mu_n}{n!}
 + {o}(|k|^{a}),
\ee
where the prime on the summation sign indicates that $n$ takes only
even values. The remainder is of order ${o}(|k|^{a})$ as $k \to 0$. In
this finite sum we only know the exponent $a$ and the moments $\mu_0=1
$ and $ \mu_2=1/4$. Now we calculate the unknown finite moments of the
scaling form, $\mu_n$ with $2< n \leq [a]$. This is done by insert
ing \Ref{A4} into the kinetic equation \Ref{phi-scale-eq}, yielding
the recursion relation,
\be \Label{A5}
\mu_n = (1/\gamma_n)  \sum_{l=2}^{n-2} h(l,n-l) \mu_l \mu_{n-l}
\ee
with initialization $\mu_2=1/4$, where $l$ and $n$ are even. The
solutions $\mu_n$ for $n=4,6,8$ are shown in Fig.~6 as a function of
the coefficient of restitution $\alpha$. The physical branches of
these functions are the ones that start positive at $\alpha =1$.
Furthermore we observe that the root $s=a$ of the transcendental
equation \Ref{trans-eq}, $\gamma_s =\lambda_s-\half s\lambda_2=0$,
indicates that   $\gamma_s$ changes sign at $s_1=a$ (see open circles
in Fig.4). This change of sign, where the branch becomes unphysical,
can according to Section 6 be interpreted as all moments $\mu_n$ with
$n>s_1=a$ ($n$ on unstable branch) becoming {\em divergent}. That this
interpretation is the correct one, has already been demonstrated in
the Section 7.1, where it is shown that for $n>a$ the reduced moments
$\mu_n(t) \to \infty $ as $t \to \infty$.

The recursion relation \Ref{A5} for the moments $\mu_n$ in the
one-dimensional case is again a bit pathological in the sense that the
{\em stable} branch $(s_0=2 < s \leq s_1=3)$ contains only one single
integer label, i.e. $s=3$. So only $\mu_0=1$ and $\mu_2=1/4$ are
finite, and all other moments are infinite, in agreement with the
exact solution of Baldassarri et al.

The recursion relation \Ref{A5} has a {\em second} set of solutions
$\{ \mu_n \}$, simply defined by iterating the recursion relation for
$n$ arbitrary large. This set contains  {\em negative} moments $\mu_n$
\cite{Bobyl-00}. The argument is simple. Consider $\mu_n$ in \Ref{A5}
with $n=[a]+1$. Then the pre-factor $1/\gamma_{[a]+1}$ on the right
hand side of this equation is negative because the label $[a] +1
> a $ is on the unstable branch of the eigenvalue spectrum in Fig.4,
while all other factors are positive. This implies that the
corresponding scaling form $f (c)$ has negative parts, and is
therefore physically not acceptable. We also note that the moments
$\mu_n$ of the physical and the unphysical scaling solution $\phi(k)$
coincide as long as both are {\it finite} and {\it positive} in the
$\alpha-$interval that includes $\alpha =1$. These unphysical
solutions are also shown in Fig.6.

\setcounter{section}{7}
\setcounter{subsection}{0}\setcounter{equation}{0}
\section{Driving and non-equilibrium steady states}
\subsection{Energy balance}
The present section is devoted to the study of systems of inelastic
Maxwell particles with energy input. Here the system may or may not be
able to reach a non-equilibrium steady state (NESS). To reach a
spatially homogeneous steady state, energy has to be supplied
homogeneously in space. This may be done by applying an external
stochastic force to the particles in the system, or by connecting the
system to a thermostat, modelled by a frictional force with {\it
negative} friction. Complex fluids (e.g. granular) subject to such
forces can be described by the microscopic equations of motion for the
particles, $\dot{\br}_i =\bv_i$, and $\dot{\bv}_i = {\bfa}_i +
\tilde{{\bf \xi}}_i$ $(i=1,2, \cdots)$, where ${\bfa}_i$ and
$\tilde{{\bf \xi}}_i$ are the possible friction and random forces
respectively. If needed one may also include  in ${\bfa}_i$ a
conservative (velocity independent) force.

Regarding the Negative Friction (NF) thermostats, the most important
and most common choice \cite{MS00} is a friction, linear in the
velocity, $+\gamma {\bv}$, the so-called {\it iso-kinetic} or {\it
Gaussian thermostat} \cite{evans-book,Sinai-isokin}. A second example
of a negative friction force is ${\bfa} = \zeta \hat{{\bv}}$, which is
acting in the direction of the particle's velocity, but independent of
its speed. Furthermore, the external stochastic force, $\tilde{{\bf
\xi}}_i$,  is taken to be Gaussian white noise with zero mean, and
variance,
\be \Label{WN-strength}
\overline{ \tilde{\xi}_{i,\alpha}(t) \tilde{\xi}_{j,\beta}(t')} = 2D
\delta_{ij}\delta_{\alpha\beta} \delta(t-t'),
\ee
where $\alpha,\beta$ denote Cartesian components, and $D$ is the noise
strength.  The Boltzmann equation, describing a spatially homogeneous
system driven in this manner, takes the form,
 \be \Label{BE-driven}
\partial_t F(v)+ ({ \mbox{$\bnabla_v$}} \cdot {\bfa}  - D \nabla_v^2) F(v)=I(v|F),
\ee
where ${\cal F}= \mbox{$\bnabla$}_v \cdot \bfa - D \nabla_v^2$
represents the driving term.

Next we consider the {\it balance equation} for the granular
temperature in driven cases, where the external input of energy
counterbalances the collisional cooling, and may lead to a NESS. We
proceed in the same manner as for the free case, and apply $(\int d\bv
v^2)$ to the Boltzmann equation in \Ref{BE-driven} with the result,
\be \Label{T-balance}
 d \av{v^2}/dt = \int d{\bv}\, v^2 I(v|F) + 2 \av{\bv \cdot {\bfa}}
+2dD.
\ee
The second and third term  are obtained from the driving term in
\Ref{BE-driven} by performing partial integrations. The most common
way of driving dissipative fluids in theoretical and MD studies
\cite{vNE98,MS00,BN-K-02,Cercig-00,ignacio-I,peng-ohta} is by Gaussian
white noise (WN) $({\bfa}=0; D \neq 0)$. We include in our studies
also the two types of NF-thermostat $({\bfa} \neq 0; D= 0)$, discussed
above. The resulting energy balance equation is,
\be \Label{vo-driven}
\frac{d v_0}{dt} = \left\{
\begin{array}{ll}
\frac{2D}{v_0}- \frac{pq}{d} v_0  & \quad (\mbox{WN})\\[1.5mm]
\frac{2\zeta \langle |c| \rangle }{d} - \frac{pq}{d} v_0 & \quad
\mbox{(const NF)}\\[1.5mm]
(\gamma - \frac{pq}{d}) v_0 & \quad
(\mbox{iso-kin})\\
\end{array} \right.
\ee
Here we have used the relation, $\langle |v| \rangle = v_0 \langle |c|
\rangle$, where the last average $\langle |c| \rangle$ is a moment of
$f(c)$. One sees that the collisional loss is counterbalanced by the
heat, generated by randomly kicking the particles or by the negative
friction of the Gaussian thermostat.

The stationary solutions of the first two equations are stable
attractive fixed points, which are approached at an exponential rate,
\be \Label{vo-NESS}
v_0(\infty) = \left\{
\begin{array}{cl}
\sqrt{2dD/pq} & \quad (\mbox{WN})\\[1.5mm]
\frac{2 \zeta \langle |c| \rangle}{pq}   & \quad (\mbox{const NF})\\
\end{array} \right.
\ee
The case of driving by an iso-kinetic thermostat (with {\it linear}
negative friction) is a marginal case, discussed in \cite{orsay},
because stationarity is only reached when the friction constant has
exactly the value $\gamma = pq/d$, in which case any initial value
$v_0(0)$ is stationary. So, for an inelastic Maxwell gas, driven by an
iso-kinetic thermostat, there does {\it not exist} a stationary state,
because the general solution of the rate equation for arbitrary value
of $\gamma$ is,
\be \Label{FC-margin}
v_0(t)= v_0(0) \exp[(\gamma -pq/d)t] \qquad (\mbox{iso-kin}),
\ee
which may be increasing or decreasing as $t \to \infty$, depending on
the inequalities, $\gamma<\gamma_0$ or $\gamma > \gamma_0$.

\subsection{Scaling equation}

The equations \Ref{BE-driven} and \Ref{vo-NESS} show that the NESS
solution $F(v,\infty)$ depends strong\-ly on the mode of energy supply.
To exhibit possible universal features of the solution we measure
velocities in their typical magnitude, i.e.~the r.m.s.~velocity
$v_0(\infty)$, just like in thermal equilibrium, and introduce the
rescaled distribution,
\be \Label{BE-f-scale}
F(v,\infty) = \frac{1}{v^d_0(\infty)} f\left(\frac{v}{v_0(\infty)}
\right).
\ee
The integral equation for the scaling form $f(c)$ follows in this case
 by inserting \Ref{BE-f-scale} in \Ref{BE-driven}, and setting $\partial_t
F=0$, i.e.
\ba\Label{eq-f-driven}
I(c|f) = \left\{ \begin{array}{clll} \displaystyle
-\frac{D}{v_0^{2}(\infty)} \nabla_{c}^2 f &=& -\displaystyle
\frac{pq}{2d}
\nabla_{c}^2 f & \qquad \mbox{(WN)}\\
\displaystyle \frac{\zeta}{v_0 (\infty)} \bnabla_{c}\cdot(\hat{\bc} f)
&=& \displaystyle \frac{pq}{2 d \:\langle |c| \rangle }
\bnabla_{c}\cdot (\bhat{\bc} f) & \qquad \mbox{(const NF)}
\end{array}\right.
\ea
In the second equality on both lines $v_0(\infty)$ has been eliminated
with the help of \Ref{vo-NESS}.

Next we consider the special case of a system driven by an iso-kinetic
thermostat $(\bfa = \gamma \bv ; D=0)$, where $F(v,t) $ does not
approach a NESS, but rapidly reaches a scaling state, described by
\Ref{sim-sol} and having the time dependent r.m.s. $v_0(t)$ in
\Ref{FC-margin}. In that case the terms $\partial_t F$ and ${\cal F}F$
 in Eq. \Ref{BE-driven} produce respectively the terms
$( -\gamma +pq/d )\:\bnabla_c \cdot (\bc f)$ and $\gamma\bnabla_c
\cdot (\bc f)$. So the terms containing the friction constant $\gamma$
cancel. The scaling equation for the iso-kinetic thermostat then
becomes,
\be \Label{eq-f-driven-FC}
I(c|f) = \frac{pq}{d} \bnabla_c \cdot (\bc f) \qquad (\mbox{iso-kin}).
\ee
This scaling equation is {\it identical} to the one derived in
\Ref{FC-scale-BE} for free cooling, and no trace of the friction
constant $\gamma$ of the iso-kinetic thermostat remains. For the case
of inelastic hard spheres the equivalence of the integral equations
for the scaling form in both cases has been observed before by
Montanero and Santos \cite{MS00}. However the big difference between
inelastic Maxwell particles and inelastic hard spheres is that the
latter system has an energy balance equation with a stable attracting
fixed point solution, but a NESS does not exist in the former case.

\subsection{ High energy tails}

The scaling equations for inelastic Maxwell particles
\cite{ME+RB-rapid} in the previous subsection cannot be solved
exactly. However its asymptotic solution for $ c \gg 1$ can be
determined by the same procedure as used  for inelastic hard spheres
\cite{vNE98}.  To do so we neglect the gain term in $I(c|f)$ in
\Ref{eq-f-driven}. Then $I(c|f)$ is replaced by $I_{\mbox{loss}}(c|f)
\sim  -  f(c)$, and the asymptotic solutions of \Ref{eq-f-driven} are
found in the form of stretched Gaussians, $f(c) \sim \exp[-\beta c^b]$
with positive $b$ and $\beta$. One then verifies { \it a posteriori}
that the stretched Gaussians are indeed consistent solutions of
\Ref{eq-f-driven} and \Ref{eq-f-driven-FC} by  substitution them back
into $I(c|f)$, and showing \cite{ME+RB-rapid} that the loss term is
asymptotically dominant over the gain term as long as exponent $b$ and
coefficient $\beta $ are positive.

After these preparations we insert the stretched exponential form into
the Boltzmann equation \Ref{eq-f-driven}, and match the leading
exponents on both sides of the equation, as well as the coefficients
in the exponents of these terms. This gives the following results for
the asymptotic high energy tail,  ${f}(c) \sim \exp[-\beta c^b]$, in
$d-$di\-men\-sional inelastic Maxwell models,
\be \Label{tail-maxwell}
\begin{array}{lll}
b=1 &\quad \beta =\sqrt{\frac{2d}{pq}}
& \quad \mbox{(WN)} \\[2mm]
b=1 &\quad
\beta= \frac{2d \langle |c|\rangle}{pq}
& \quad\mbox{(const NF)}  \\[2mm]
b=0 & \quad\mbox{inconsistent} & \quad\mbox{(iso-kin)}.
\end{array}
\ee

We conclude this subsection  about driven inelastic Maxwell models
with some comments:

Why is the result, $b=0$, inconsistent for the {\it iso-kinetic}
thermostat in this model? Taking the limit $ b \to 0$ in $\exp[-\beta
c^b]$ suggests that $f(c)$ has indeed a power law tail. As we have
seen in \Ref{eq-f-driven-FC}, the scaling equation for the IMM driven
by this thermostat is equivalent with the scaling equation for free
cooling, and we know from the analysis in Section 6.2 that $f(c) \sim
1/c^{a+d}$ has indeed a power law tail. However the exponent $a$ that
would have been obtained from \Ref{eq-f-driven-FC} with $I$ replaced
by $I_{\mbox{loss}} \sim -f(c)$ does not yield a consistent solution.
In the limit $b \to 0$ the gain term $I_{\mbox{gain}}$  can no longer
be neglected with respect to the loss term. That this is indeed the
case can be seen from \Ref{trans-eq}, where the terms $\eta^a_+$ and $
\eta^a_-$ originate from the gain term. These terms give substantial
contributions to the value of $a$, and are even dominant for small
values of $\alpha$. We also note that in the case of driving by an
iso-kinetic thermostat---which turns out to be equivalent to the free
cooling IMM system---the driven system does not reach a nonequilibrium
steady state, but keeps either heating up or cooling down, depending
on the thermostat strength $\gamma$.

{\it White noise} driving and positive $b$ lead for all $d\geq 1$ to
consistent asymptotic solutions of the scaling equations with
overpopulated high energy tails of simple exponential type, $f(c)\sim
\exp[-c\sqrt{2d/pq}]$ (Refs.\cite{ME+RB-rapid,BC-WNtail}),
in complete agreement with the asymptotic
result \Ref{c6}, and in qualitative agreement with the corresponding
result \Ref{h11} for BGK-models. Here {\it all} moments $\int d\bc c^n
f(c) <\infty$. This would not be the case for power law tails.  The
one-dimensional version of this problem has been extensively studied
 by Nienhuis and van der Hart \cite{Nienhuis}, and by Antal et al. \cite{Droz}
using MC simulations of the Boltzmann equation. MC simulations of the
two-dimensional version of this problem have been carried out in Ref.
\cite{orsay}.

When the IMM system is driven by a {\it constant NF-}thermostat the
scaling function also shows an exponential tail, $f(c) \sim \exp[ -2d
\langle c \rangle c /pq]$, with a coefficient that depends on the
first moment $\langle c \rangle$ of the complete scaling $f(c)$. In
Refs. \cite{vNE98,EB-Springer,orsay} methods have been developed to
calculate these moments perturbatively.

For comparison we also quote the high energy tail for $d-$dimensional
inelastic hard spheres, which are also of the form of stretched
Gaussians, and we quote the results for the exponents $b$ and the
coefficients $\beta$ for the different modes of energy supply, i.e.
\be \Label{tail-IHS}
\begin{array}{lll}
b=3/2 &\quad \beta =\sqrt{\frac{2d \beta_1}{pq \kappa_3}}
& \quad \mbox{(WN)} \\[2mm]
 b=2 &\quad
\beta= \frac{\beta_1 \langle |c|\rangle}{pq \kappa_3} \simeq
\frac{1}{\sqrt{2} pq}
& \quad\mbox{(const NF)}  \\[2mm]
b=1 & \quad \frac{d \beta_1}{pq \kappa_3} & \quad\mbox{(iso-kin)}
\end{array}
\ee
where $\beta_1$ is given in \Ref{beta-s}, $\langle c \rangle = \int
d\bc c f(c)$ and $\kappa_3 = \int_n \int d\bc d\bfu |(\bc -\bfu)\cdot
\bn|^3 f(c)f(u)$ \cite{EB-Springer}. For $\alpha $ close to unity the
replacement of $f(c)$ by the Gaussian $ \pi^{-d/2}\exp[-c^2]$ gives a
good approximation, yielding $\langle c \rangle \simeq 1 /[
\sqrt{\pi}\beta_1]$ and $\kappa_3 = \sqrt{2/\pi}$. The results for the
WN- and iso-kinetic thermostat have been first derived in
\cite{vNE98}, and confirmed by MC simulations in \cite{MS00}. The
theoretical result for the {\it constant} NF thermostat was first
derived in \cite{MS00}, but its consistency was questioned. For that
reason we have verified the consistency of the result \Ref{tail-IHS} a
posteriori, and we confirm that it is indeed an asymptotic solution of
\Ref{eq-f-driven} with the full Boltzmann collision operator, as long
as $\alpha <1$.

\setcounter{section}{8}
\setcounter{subsection}{0}\setcounter{equation}{0}
\section{Conclusions}

In the present paper we have reviewed the new developments on
anomalous velocity distributions in gases of inelastic Maxwell models
(IMM), and compared the results with those for the proto-typical
granular model, the inelastic hard spheres (IHS). The velocity
distributions, obtained in this article, are scaling or similarity
solutions \Ref{sim-sol} of the nonlinear Boltzmann equation.

The dominant common feature in all these results is that the
inelasticity of the collisions creates {\it over-populations} of high
energy tails of the velocity distribution $F(v,t)$, when compared to
the Gaussian Maxwell-Boltzmann distribution $F(v,\infty)$ in thermal
equilibrium of systems of  elastic particles. At finite inelasticity
$(\alpha <1)$ the overpopulations in the high energy tails are power
laws, $f(c) \sim 1/c^{a+d}$ (free cooling in IMM), or stretched
Gaussians, $\exp[-\beta c^b]$ with $0< b < 2$ (free cooling IHS and
driven IMM and IHS).

An intermezzo, presented in Section 4, shows that the results obtained
for inelastic Maxwell models are rather robust. We consider an
extremely simplified inelastic BGK-model, and  show that the resulting
over-populations of high energy tails, both with and without energy
input, are qualitatively the same as for the nonlinear Boltzmann
equation of inelastic Maxwell models in $d-$dimensions.

Returning to the main menu of IMM's, we note that the degree of
overpopulation  is decreasing $(b \uparrow 2)$  with the increasing
efficiency to randomize the velocities of the highly energetic
particles, either by collisions or by the external forcing terms. As
the IHS's have a larger collision frequency, $\propto g$, than the
IMM's with a constant collision frequency, the IHS have smaller
over-populations than the IMM's. Because external white noise applied
to freely cooling inelastic gases adds an extra mechanism  for
randomization, the tails in the driven cases show lower
over-populations than in the freely cooling case.

A intriguing question about over-populated tails: "Are power laws or
stretched exponentials the generic form of over-populated tails in
inelastic models", is difficult to answer because we have only
information from two different interaction models. In two very recent
articles \cite{EB-Springer,orsay} new classes of inelastic models have
been introduced, that correspond to soft spheres with repulsive
interactions $(1/r^s)$. These soft spheres have a collision frequency
$ \propto g^\nu$ with $\nu = 1-2(d-1)/s$. The limit $s \to \infty$, or
$\nu \uparrow 1$, corresponds to strong IHS interactions at high
energy. In the limit $\nu \downarrow 0$, analogously to $s \downarrow
2(d-1)$, the interactions decrease. For $\nu \geq 0$  the
IMM-interactions $(\nu=0)$ are the weakest of all. The results for
these inelastic soft spheres models \cite{EB-Springer,orsay},
corresponding to \Ref{tail-maxwell} and \Ref{tail-IHS}, are again
stretched Gaussians $f(c) \sim \exp[-\beta c^b]$ with
\be
\begin{array}{ll}
b=\half (\nu +2) & \quad \mbox{(WN; $\:\nu \geq 0 $)}
\\[2mm]
b=\nu & \quad \mbox{(iso-kin; $\nu>0$  )}\\
\end{array}
\ee
The limit $b = \nu \downarrow 0$ for the iso-kinetic thermostat, which
also corresponds to free cooling (see Section~8), is consistent with a
power law tail. This analysis shows that the generic type of
over-population is a stretched Gaussian. A freely cooling IMM with
$f(c) \sim 1/c^{a+d}$ is an isolated borderline case, that is most
likely not the best model to describe the short range, hard core
impulsive interactions of granular particles.

At the end of Section 3 we have seen in Table I the phenomenon of
compressed Gaussians with $b>2$, corresponding to under-population of
high energy tails. The solutions with two delta peaks are extreme
cases of compressed Gaussians. More extensive discussions of
compressed Gaussians and  of peak-splitting in  one-dimensional
inelastic models can be found in  \cite{BBRTW,SE-condmat,orsay}.

\section*{Acknowledgements}
The authors want to thank J.W. Dufty, A. Santos, E. Trizac, A. Barrat
and K. Shundyak for helpful discussions and correspondence. This work
is supported by DGES (Spain), Grant No BFM-2001-0291.

\end{document}